\documentclass[11pt]{article}
\usepackage{amssymb}
\usepackage{amsfonts}
\usepackage{amsmath}
\usepackage{graphicx}
\usepackage{latexsym}

\newcommand{\bb}{\begin{eqnarray}}
\newcommand{\ee}{\end{eqnarray}}
\newcommand{\beq}{\begin{equation}}
\newcommand{\eeq}{\end{equation}}
\newcommand{\ba}{\begin{array}}
\newcommand{\ea}{\end{array}}

\title{Multicomponent integrable wave equations II. \\
Soliton solutions}

\author{Antonio Degasperis\\
Dipartimento di Fisica, Universit\`a di Roma ``La Sapienza'', \\
and Istituto Nazionale di Fisica Nucleare, Sezione di Roma, Rome, Italy. \\
E-mail: antonio.degasperis@roma1.infn.it
\and
Sara Lombardo\\
Department of Mathematics, Faculty of Sciences\\
Vrije Universiteit, De Boelelaan 1081a, 1081 HV Amsterdam, The Netherlands\\
Email: sara@few.vu.nl\\
School of Mathematics, Alan Turing Building \\
University of Manchester, Upper Brook Street, Manchester M13 9EP, UK\\
Email: sara.lombardo@manchester.ac.uk}
\date{}

\begin{document}

% the next command turns on and off the option of numbers for each line
%\linenumbers  

\maketitle

\begin{abstract}
The Darboux--Dressing Transformations developed in \cite{ddt} are here applied to construct soliton solutions for a class of boomeronic--type equations. The vacuum (i.e. vanishing) solution and the generic plane wave solution are both dressed to yield one soliton solutions. The formulae are specialised to the particularly interesting case of the resonant interaction of three waves, a well-known model which is of boomeronic--type. For this equation a novel solution which describes three locked dark pulses (simulton) is introduced. 

\vspace{0.2cm}

\end{abstract}

\newpage

\section{Introduction}
 Many phenomena of wave propagation  in one dimensional media  in various physical contexts are modelled by systems of Partial Differential Equations (PDEs). Among them, integrable models are of special interest and are the focus of our investigation. In fact the present article follows our work  \cite{ddt} to which we refer the reader for the general background and motivations. Our interest in this subject is particularly motivated by the introduction and investigation of  boomeron solutions in models of interest in nonlinear optics. These solutions are soliton solutions whose behavior differs from  that of standard solitons (such as for the KdV and NLS equations). In fact they have different asymptotic velocities at $t=+\infty$ and $t=-\infty$. They were first introduced long time ago in \cite{CD1976} (see also \cite{D1978}) in connection with the hierarchy of the matrix Korteweg--de Vries equation, and more recently boomerons appeared again in geometry \cite{DRS2002} and optics \cite{dcbw}, \cite{cbdw2006}, \cite{cbdw2007}, \cite{cbdw2008}, \cite{mantsyzov}.
The special properties of boomerons stem from the fact that they are solutions of multicomponent field equations whose integrability requires a matrix formulation of the corresponding Lax pair (see below) together with a richer (non commutative) algebraic structure. The characteristic property of this structure is that the flows of the attached hierarchy may not commute with each others (see \cite{CD2004} and \cite{CD2006}).
The technique of constructing  soliton solutions we adopt here goes via the  Darboux--Dressing Transformations (DDT) which yield  a new (\emph{dressed}) solution from a given (\emph{naked} or \emph{seed}) solution of the system of wave equations of interest. This method may be formulated as a change of  the Lax pair of linear equations via a transformation which adds one pole to the dependence of its solution on the complex  spectral variable. Though the formal setting of this construction is well known (\cite{DDT1}--\cite{DDT4}), the actual construction is detailed in  \cite{ddt}  where the explicit algorithmic procedure is given. In particular, there we show that the position of the pole, whether it lies off or on the real axis, leads to different expressions of the dressed solution. In its turn, as we show here, this difference plays a crucial role in accounting for different boundary values of the wave fields, namely boundary values which are appropriate to bright solitons (localised pulses with exponential decay) or dark solitons (localised pulses in a non vanishing background).  
\newline
To the aim of making this article as much self--contained as possible, in this first section we set up the general formalism  of matrix evolution equations we have introduced in  \cite{ddt} by keeping the  same notation as introduced there. In Section 2 the DDT construction, which is discussed in details in \cite{ddt}, is tersely reported. Section 3 is devoted to the soliton expressions of both  bright and dark type in the general matrix form. However, as for the investigation of these expressions, we prefer to focus on one particular reduction. This and other reductions lead to models which seem to deserve special attention because of their potential applicability. Thus Section 4 is devoted to novel simulton solutions of the well know system  which describes the interaction of three waves in quadratic media with the resonant condition $\omega_2-\omega_1=\omega_3$ and $k_2-k_1=k_3$ on their frequencies and wave numbers, respectively. 
%The second reduction, which is introduced in Section 5, is a special case of five waves which interact with each other as two resonant triads satisfying the frequency and wave number  conditions $\omega_1+\omega_2=\nu_1+\nu_2=\omega_3$ and $k_1+k_2=q_1+q_2=k_3$. 
The last Section 5 contains few remarks and conclusions.
\newline
\noindent
The starting point of our investigation is the following Lax pair of matrix Ordinary Differential Equations (ODEs):
\begin{equation}\label{lax} 
\psi_x=X\psi\,, \quad\psi_t=T\psi\,, 
\end{equation}
where $\psi$, $X$ and $T$ are $N\times
N$ square matrices, $\psi=\psi(x,t,k)$ being a common
solution of the two linear ODEs
(\ref{lax}), while $X=X(x,t,k)$ and $T=T(x,t,k)$ depend on the coordinate $x$, the time $t$ and the complex spectral parameter $k$ according to the definitions 
\begin{subequations}
\begin{equation}\label{xLaxoperator}
\begin{array}{l}
X(x,t,k) =ik\sigma+Q(x,t)\,, 
 \end{array} 
\end{equation}
\begin{equation}\label{tLaxoperator}
\begin{array}{l}
T(x,t,k)=2ikC -\sigma W+\sigma [C,Q(x,t)]\, ,
 \end{array} 
\end{equation}
\end{subequations}
where $[A,B]$ stands for the commutator $AB-BA$.
We notice that, in order to semplify our analysis, we have chosen the $X$ and $T$ matrices to be first degree in the spectral variable $k$. This choice has two consequences, namely it lets the two independent variables $x$ and $t$ play a similar role so that their interpretation as \emph{space} and \emph{time} can change according to the physical application, and, secondly, it leads to wave equations which are dispersionless when linearized around the vanishing solution. Dispersion terms can be easily introduced with only technical changes of the method (see  \cite{ddt} where the dispersive term of Schr\"odinger type has been considered). In the equations  (\ref{lax}) $\sigma$ is the diagonal constant matrix
\begin{equation}\label{sigma} 
\sigma=\left( \begin{array}{cc} \mathbf{1}_{N^{(+)}\times N^{(+)}} & 
\mathbf{0}_{N^{(+)}\times N^{(-)}} \\ \mathbf{0}_{N^{(-)}\times
N^{(+)}} & -\mathbf{1}_{N^{(-)}\times N^{(-)}} \end{array} \right)\,,
\end{equation}
while $C$ is an arbitrary constant block--diagonal
matrix,
\begin{equation}\label{Cmatrices}
C=
\begin{pmatrix}
C^{(+)} & \mathbf{0}_{N^{(+)}\times N^{(-)}} \\ 
\mathbf{0}_{N^{(-)}\times N^{(+)}} & C^{(-)}
\end{pmatrix}\,\,, 
\end{equation}
where $C^{(+)}$, respectively $C^{(-)}$, are $N^{(+)}\times
N^{(+)},$ respectively $N^{(-)}\times N^{(-)},$ constant square
matrices, with $N=N^{(+)} + N^{(-)}$, $N^{(+)}$ and $N^{(-)}$ being arbitrary positive integers. The \emph{potentials} $Q=Q(x,t)$ and $W=W(x,t)$ are the off-diagonal and, respectively, diagonal block matrices 
(a superimposed dagger stands for Hermitian conjugation)
\begin{equation}\label{qwblock}
Q=\left( \begin{array}{cc} \mathbf{0}_{N^{(+)}\times N^{(+)}} & S^{(+)}U^{\dagger}S^{(-)}
\\ U & \mathbf{0}_{N^{(-)}\times N^{(-)}} \end{array} \right) \,,\quad
W=\left( \begin{array}{cc}  W^{(+)} & \mathbf{0}_{N^{(+)}\times N^{(-)}}\\ 
\mathbf{0}_{N^{(-)}\times N^{(+)}} & W^{(-)} \end{array} \right) \,,
\end{equation}
where the block $U=U(x,t)$ is a  rectangular $N^{(-)}\times N^{(+)}$ matrix while the two  blocks $W^{(\pm)} $
are square $N^{(\pm)}\times N^{(\pm)}$ matrices. Moreover 
 $S^{(+)}$ and  $S^{(-)}$ are, respectively, 
$N^{(+)}\times N^{(+)}$ and $N^{(-)}\times N^{(-)}$ diagonal matrices
whose diagonal elements $s^{(\pm)}_n$, with no loss of generality, are just signs, namely
\begin{equation}\label{signs}
S^{(\pm)}=\mbox{diag}\,\{s^{(\pm)}_1,\cdots,s^{(\pm)}_{N^{(\pm)}}\}\,,\quad
{s^{(\pm)}_{n}}^2=1\,\,\,.
\end{equation}
The compatibility of the two ODEs (\ref{lax}) entails the matrix first order differential equations
\begin{subequations}
\label{redboomeron}
\begin{equation}\label{Uboomeron}
\begin{array}{l}
U_{t}=U_{x}\,C^{(+)} - C^{(-)}\,U_{x} + U\,W^{(+)} + W^{(-)}\,U
\end{array}
\end{equation}
\begin{equation}\label{Wplus}
W_{x}^{(+)}=\left[ C^{(+)},\,S^{(+)}\,U^{\dagger }\,S^{(-)}\,U\right] ,
\end{equation}
\begin{equation}\label{Wminus}
W_{x}^{(-)}=\left[ C^{(-)},U\,S^{(+)}\,U^{\dagger }\,S^{(-)}\right] ,
\end{equation}
\end{subequations}
provided the square $N^{(\pm)}\times N^{(\pm)}$ matrices $C^{(\pm )}$ and $W^{(\pm)} $ satisfy
the conditions 
\begin{equation}\label{Cond1}
C^{(\pm)\dagger}=\,S^{(\pm)}\,C^{(\pm)}\,S^{(\pm)}\,
\end{equation}
and 
\begin{equation}\label{CondW}
W^{(\pm)\dagger}=\,-\,S^{(\pm)}\,W^{(\pm)}\,S^{(\pm)}\,\,\,. 
\end{equation}
The reduced form (\ref{qwblock}) of the matrix $Q(x,t)$ is well motivated by the fact that it captures several interesting models of multicomponent wave interactions in weakly nonlinear media, see \cite{CD2004},  \cite{CD2006} and Sections 4 and 5.

%%%%%%%%%%%%%SEZIONE 2 %%%%%%%%%%%%%%%%%%%%%%%%%%%%%
\section{The DDT algorithm}\label{sec:ddtalgo}
The method of construction of explicit solutions of the Lax equations (\ref{lax}), and therefore of the nonlinear evolution equations  (\ref{redboomeron}), may be formulated in more than one  way \cite{DDT1}--\cite{DDT4}. Here we sketch the scheme we developed in \cite{ddt} where full details are given. 
\newline
Let $Q^{(0)}(x,t)$ and $W^{(0)}(x,t)$ be a pair of matrices with the same block structure (\ref{qwblock})  of $Q(x,t)$ and $W(x,t)$. This assumption defines therefore also the block matrices $U^{(0)}(x,t)$ and $W^{(0)(\pm)}(x,t)$. Let then $\psi^{(0)}(x,t,k)$ be a corresponding
nonsingular (i.e. with nonvanishing determinant) matrix solution of
\begin{equation}\label{0Laxpair}
\psi^{(0)}_x=X^{(0)}\psi^{(0)}\,,
\quad \psi^{(0)}_t=T^{(0)}\psi^{(0)}
\end{equation}
where $X^{(0)}(x,t,k)$ and $T ^{(0)}(x,t,k)$ have the expressions (\ref{xLaxoperator}) and (\ref{tLaxoperator}) whit $Q$ and $W$ replaced by $Q^{(0)}$ and $W^{(0)}$. 
Since both compatibility conditions,
$\psi^{(0)}_{xt}=\psi^{(0)}_{tx}$ and $\psi_{xt}=\psi_{tx}$ are
satisfied, $U^{(0)}(x,t), W^{(0)(\pm)}(x,t)$ and $U(x,t), W^{(\pm)}(x,t)$ are just two different solutions of the same matrix evolution equations (\ref{redboomeron}). Consider now the matrix $D=\psi \psi^{(0)-1}$ or, equivalently, the relation 
\begin{equation}\label{DDT}
\psi(x,t,k)=D(x,t,k)\psi^{(0)}(x,t,k)\,\,, 
\end{equation}
which can be viewed as a transformation  $\psi^{(0)} \rightarrow \psi$. The DDT method consists in searching for the dressing matrix $D(x,t,k)$ for a given $\psi^{(0)}$.
The  expression of the matrix $D(x,t,k)$ is found by assigning \emph{a priori} its explicit dependence on the spectral parameter $k$. We assume that this transformation is  near identity and adds one simple pole to the $k-$dependence of the solution $\psi^{(0)}(x,t,k)$, namely
\begin{equation}\label{elementaryDDT}
D(x,t,k)=\mathbf{1}+\frac{R(x,t)}{k-\alpha}\,,
\end{equation} 
where $R(x,t)$ is the residue matrix and the parameter $\alpha$ is the pole position in the complex plane of the spectral variable $k$. In \cite{ddt} we derive the expression of the residue matrix $R(x,t)$ and we show that this expression crucially depends on the position of the pole $k=\alpha$, the distinction being between the case in which the pole is off  the real axis, $\alpha\neq \alpha^*$, and the case in which the pole is on the real axis, $\alpha = \alpha^*$. In the next section we show that this distinction is relevant when constructing soliton solutions with vanishing or non vanishing asymptotic values of $U(x,t)$ as $|x|\rightarrow \infty$. Here we merely report the following expression of the new solution $U(x,t)$, $W^{(\pm)}(x,t)$ in these two cases:
\begin{subequations}\label{realBacklundcompl}
\begin{equation}\label{realUBacklundcompl}
U=U^{(0)}-\frac{z^{(-)}z^{(+)}
{}^\dagger S^{(+)}}{\mathcal{E}(x,t)}\,,
\end{equation}
\begin{equation}\label{realwpBacklundcompl}
W^{(\pm)}=W^{(0)(\pm)}- \frac{\left[C^{(\pm)}\,,\,z^{(\pm)} z^{(\pm)}{}^\dagger S^{(\pm)}\right]}{\mathcal{E}(x,t)}\,\,.
\end{equation}
\end{subequations}
 $S^{(\pm)}$ is the sign matrix (\ref{signs}), $C^{(\pm)}$ is the $N^{(\pm)} \times N^{(\pm)}$ constant matrix as defined in (\ref{Cmatrices}) with the condition (\ref{Cond1}) and the $N^{(\pm)}$--dimensional vector $z^{(\pm)}$ is defined below, see (\ref{Czsolution}). The difference between the two cases is mainly in the expression of the real denominator function $\mathcal{E}(x,t)=\mathcal{E}^*(x,t)$. Thus when the pole $\alpha$ is complex with non vanishing imaginary part (i.e. $\alpha\neq \alpha^*$), the denominator becomes \cite{ddt}
\begin{equation}\label{Cdenom}
\mathcal{E}(x,t)=\mathcal{E}_C(x,t)\equiv  \frac{i}{2(\alpha - \alpha^*)} [ (z^{(+)},S^{(+)}z^{(+)})-(z^{(-)},S^{(-)}z^{(-)})] \,\,\,,
\end{equation}
while, if the pole $\alpha$ is real (i.e. $\alpha= \alpha^*$), the expression of the denominator is quite different from the previous one and reads \cite{ddt}
\begin{equation}\label{Rdenom}
\mathcal{E}(x,t)=\mathcal{E}_R(x,t)\equiv \mathcal{E}_0+\frac{1}{2} \int_{x_0}^xdx' g(x',t)+ \int_{t_0}^tdt' m(x_0,t')
\end{equation}
where $\mathcal{E}_0\,,\,x_0$ and $t_0$  are  arbitrary real constants and the scalar functions  $g(x,t)$ and $m(x,t)$ are defined by the expressions
\begin{equation}\label{fghmcomp}
\begin{array}{l}
g(x,t)= (z^{(+)},S^{(+)}z^{(+)})+(z^{(-)},S^{(-)}z^{(-)})\,,     \\
m(x,t)= (z^{(+)},S^{(+)}C^{(+)}z^{(+)})-(z^{(-)},S^{(-)} C^{(-)}z^{(-)})\,.    
\end{array}
\end{equation}
Moreover, in this second case, i.e. $\alpha=\alpha^*$, the two vectors $z^{(+)}$ and $z^{(-)}$ are constrained by the relation
$(z^{(+)},S^{(+)}z^{(+)})-(z^{(-)},S^{(-)}z^{(-)})=0\,$, which is a consequence of the reduction conditions (\ref{qwblock}), (\ref{Cond1}) and (\ref{CondW}) (see \cite{ddt} for details).  The bracket $(a,b)$ in the previous formulae stands for the (nonsymmetrical) scalar product of two vectors,
\begin{equation}\label{scalarprod}
(a,b)=a^\dagger b=\sum_j a_j^* b_j\,\,\,,
\end{equation}
and the vectors $z^{(+)}$ and  $z^{(-)}$ are $N^{(+)}$--dimensional and, respectively, $N^{(-)}$--dimensional. They are defined by the explicit block expression
\begin{equation}\label{Czsolution}
\left(\begin{array}{c}z^{(+)}(x,t)  \\
z^{(-)}(x,t) \end{array}\right)\,=\,\psi^{(0)}(x,t,\alpha^\ast)\,\left(\begin{array}{c}z_0^{(+)}  \\
z_0^{(-)} \end{array}\right)\,,
\end{equation}
where the two complex vectors $z_0^{(+)}$ and $z_0^{(-)}$ are constant in both cases; they are arbitrary if $\alpha\neq \alpha^*$ while, if $\alpha=\alpha^*$ is real, these constant vectors have to meet the same condition satisfied by $z^{(+)}$ and $z^{(-)}$, namely
\begin{equation}\label{zcond}
(z_0^{(+)},S^{(+)}z_0^{(+)})-(z_0^{(-)},S^{(-)}z_0^{(-)})=0\,
%\,,\,\,\mathrm{if} \,\,\alpha=\alpha^*\,
\,.
\end{equation}
As for the expression (\ref{Rdenom}), we observe that, because of the conservation law $g_t=2m_x$, the differential form $gdx+2mdt$ is exact and therefore the denominator expression (\ref{Rdenom}) can be computed as the integral
\begin{equation}\label{Rdenomin}
\mathcal{E}(x,t)=\mathcal{E}_R(x,t)= \mathcal{E}_0 + \frac{1}{2} \int_{x_0,t_0}^{x,t} [g(x',t')dx'+2m(x',t')dt'] 
\end{equation}
along any curve in the $(x,t)$ plane.

We end this section noticing that these explicit formulas  are meant to serve as the main tools to construct soliton-- and, by repeated application of DDTs, multisoliton--solutions of those systems of multicomponent wave equations which can be obtained by reduction of the general matrix PDEs (\ref{redboomeron}). 
However, they have been obtained by algebra and local integration of differential equations by leaving out two important ingredients of solutions of applicative interest, namely their boundary values and their boundedness. These two issues will be addressed in the next sections where explicit expressions of solutions will be displayed.

%%%%%%%%%%%%%%%%%%SEZIONE 3%%%%%%%%%%%%%%%%%%%%%%%%%
\section{Soliton solutions}\label{sec:solitons}
We construct now explicit solutions of the general matrix equation
(\ref{redboomeron}) by means of DDT's. The general expressions we derive here will be investigated in more details in particular cases  in the next section.   
In the present analysis, the matrix--valued function $U(x,t)$, which appears in the spectral equation $\psi_x=X\psi\,$, with (\ref{xLaxoperator}) and (\ref{qwblock}),  is restricted to be in one of  two different functional classes. The first one is the class $\mathcal{B}$ of
those (bounded) functions $U(x,t)$ which vanish sufficiently fast, for any $t$, as
$|x|\rightarrow \infty$: 
\begin{equation}\label{classB}
U(x,t)\,\in \mathcal{B}
\quad\mbox{if}\quad
\lim_{|x|\rightarrow\infty} U(x,t)=0\,.\\
\end{equation}
Bright solitons are in this class. We point out here that, if $U(x,t)$ is in $\mathcal{B}$, the matrices $W^{(\pm)}(x,t)$ which satisfy the equations (\ref{Wplus}) and (\ref{Wminus})  are necessarily  $x$--independent (possibly $t$--dependent) when $|x|\rightarrow \infty$. In particular we choose
\begin{equation}\label{WclassB}
W^{(+)}(+\infty,t)=- F^{(+)}\;,\;W^{(-)}(-\infty,t)= F^{(-)}\quad\mbox{if}\quad U(x,t)\,\in \mathcal{B}\,\,,
\end{equation}
$F^{(\pm)}$ being two arbitrary constant matrices.
 We limit our construction to the one soliton solution which obtains by choosing in the DDT algorithm the seed solution as  the trivial solution $U^{(0)}(x,t)=0, W^{(0)(\pm)}(x,t)=\mp F^{(\pm)}$. 
\newline
The second class, $\mathcal{D}$,  is the class of those (bounded) functions $U(x,t)$ which do not vanish as $|x|\rightarrow\infty$, as they  asymptotically behave like \emph{plane waves}:
\begin{equation}\label{classD1}
U(x,t)\,\,\in \mathcal{D}
\quad\mbox{if}\quad
\lim_{x\rightarrow \pm \infty} U(x,t)\,=\,U_{\pm PW}(x,t)
\end{equation}
where the functions $U_{+PW}(x,t)$ and $U_{-PW}(x,t)$ are two generally different plane wave solutions of (\ref{redboomeron}). If $U(x,t)$ is in this class, also the two matrices $W^{(\pm)}(x,t)$ which satisfy the equations (\ref{Wplus}) and (\ref{Wminus}) behave as plane waves, as shown below.  The explicit expression of plane wave solutions is reported below, but its computation requires some tedious algebra which is not reported here. In fact, it turns out that the plane wave solutions can be solutions only of two special cases of the general evolution equations (\ref{redboomeron}). The first (second) case obtains by requiring that the constant matrix $C^{(+)}$ ($C^{(-)}$) is proportional to the identity matrix with the consequence that the matrix $W^{(+)}(x,t)$ ($W^{(-)}(x,t)$) is constant, see (\ref{Wplus}) (see (\ref{Wminus})). However in this case both  $C^{(+)}$ and $W^{(+)}(x,t)$ ($C^{(-)}$ and $W^{(-)}(x,t)$) can be easily transformed away so that setting $C^{(+)}=W^{(+)}=0$ reduces the system (\ref{redboomeron})  to the two coupled equations
\begin{subequations}
\label{redboomeronminus}
\begin{equation}\label{PWUminus}
\begin{array}{l}
U_{t}= - C^{(-)}\,U_{x} + W^{(-)}\,U
\end{array}
\end{equation}
\begin{equation}\label{PWWminus}
W_{x}^{(-)}=\left[ C^{(-)}\,,\,U\,S^{(+)}\,U^{\dagger }\,S^{(-)}\right] ,
\end{equation}
\end{subequations}
 while, in the second case, namely $C^{(-)}=W^{(-)}=0$, the system (\ref{redboomeron}) reduces  to the two coupled equations
\begin{subequations}
\label{redboomeronplus}
\begin{equation}\label{PWUplus}
\begin{array}{l}
U_{t}=  U_{x}\,C^{(+)} + U\,W^{(+)}\,
\end{array}
\end{equation}
\begin{equation}\label{PWWplus}
W_{x}^{(+)}=\left[ C^{(+)}\,,\,S^{(+)}\,U^{\dagger }\,S^{(-)}\,U\right] \,\,.
\end{equation}
\end{subequations}
  Moreover, we apply a similarity transformation to  the equations (\ref{redboomeronminus}) ((\ref{redboomeronplus})) in such a way that the nonvanishing matrix $C^{(-)}$ ($C^{(+)}$)   be diagonal (and real, see (\ref{Cond1}))
\begin{equation}\label{Cmatrixminus}
C^{(-)}=\text{diag}\{c^{(-)}_1,\cdots,c^{(-)}_{N^{(-)}}\}\,\,,\,\,c^{(-)}_j=c^{(-)*}_j\,\,,
\end{equation}
in the first case, and 
\begin{equation}\label{Cmatrixplus}
C^{(+)}=\text{diag}\{c^{(+)}_1,\cdots,c^{(+)}_{N^{(+)}}\}\,\,,\,\,c^{(+)}_j=c^{(+)*}_j\,\,,
\end{equation}
in the second case. We also
 assume, for the sake of simplicity, that $c^{(\pm)}_j \neq c^{(\pm)}_n$ if $j\neq n$ and, with no loss of generality, that  the diagonal part  of the matrix $W^{(-)}$ ($W^{(+)}$) be vanishing, namely $W^{(-)}_{jj}(x,t)=0\,,\,j=1, \cdots , N^{(-)}$ $(W^{(+)}_{jj}(x,t)=0\,,\,j=1, \cdots , N^{(+)})$.
 \newline
In the first case we find that the plane wave matrix solution $U(x,t)$ has the dyadic expression \begin{equation}\label{PWUexpminus}
\begin{array}{l}
U(x,t)=\exp[i(K^{(-)}x-\Omega^{(-)} t)] A^{(-)} A^{(+) \dagger}\,\,,
\end{array}
\end{equation}
while the $N^{(-)}\times N^{(-)}$ matrix $W^{(-)}(x,t)$ is found to be
\begin{equation}\label{PWWexpminus}
\begin{array}{l}
W^{(-)}(x,t)=-i\exp[i(K^{(-)}x-\Omega^{(-)} t)] B^{(-)}\exp[-i(K^{(-)}x-\Omega^{(-)} t)]\,\,,  
\end{array}
\end{equation}
with the following specifications: $K^{(-)}$ is an arbitrary real constant diagonal  $N^{(-)}\times N^{(-)}$ matrix,
\begin{equation}\label{Kminus}
K^{(-)}=\text{diag}\{k^{(-)}_1,\cdots, k^{(-)}_{N^{(-)}}\}\,\,,
\end{equation}
with the assumption $k^{(-)}_j \neq k^{(-)}_n$ if $j\neq n$; the $N^{(-)}\times N^{(-)}$ matrix $\Omega^{(-)}$ is also a constant, real and diagonal matrix
\begin{equation}\label{Omegaminus}
\Omega^{(-)}=\text{diag}\{\omega^{(-)}_1,\cdots,\omega^{(-)}_{N^{(-)}}\}\,\,,
\end{equation}
whose entries take the expression
\begin{equation}\label{omegaminusj}
\omega^{(-)}_j=c^{(-)}_j k^{(-)}_j + (A^{(+)}\,,\,S^{(+)} A^{(+)}) \sum_{n=1,n\neq j}^{N^{(-)}} S^{(-)}_n |A_n^{(-)}|^2 \left(\frac{c^{(-)}_j-c^{(-)}_n}{k^{(-)}_j-k^{(-)}_n}\right)\,\,;
\end{equation}
$A^{(+)}$ and $A^{(-)}$ are  arbitrary constant vectors of dimension $N^{(+)}$ and, respectively,  $N^{(-)}$ and $B^{(-)}$ is the  $N^{(-)}\times N^{(-)}$ off--diagonal matrix whose  entries are
\begin{equation}\label{Bmatrixminus}
B^{(-)}_{jj}=0\,,\quad B^{(-)}_{jn}= (A^{(+)}\,,\,S^{(+)} A^{(+)})S^{(-)}_n A_j^{(-)}A_n^{(-)*} \left(\frac{c^{(-)}_j-c^{(-)}_n}{k^{(-)}_j-k^{(-)}_n}\right)\,\,,\quad j\neq n\,\,.
\end{equation}
Thus these multi--component plane wave solutions are characterized by $N^{(-)}$ wave numbers $k^{(-)}_j$ and frequencies $\omega^{(-)}_j$, $j=1,\cdots,N^{(-)}$, and by $N^{(+)}$ amplitudes $A_m^{(+)}$ and $N^{(-)}$ amplitudes $A_j^{(-)}$, which are the components of the $N^{(+)}$--dimensional vector $A^{(+)}$ and, respectively, of the $N^{(-)}$--dimensional vector $A^{(-)}$.
We complete the definition of the class  $\mathcal{D}$ of solutions  for the special case   (\ref{redboomeronminus}) of the general equation (\ref{redboomeron}) by adding the  requirement  that the asymptotic plane wave solutions $U_{-PW}$ at $x=-\infty$ and $U_{+PW}$ at $x=+\infty$, see the definition (\ref{classD1}), have the same wave numbers $k^{(-)}_j$ and frequencies $\omega^{(-)}_j$ and may differ from each other only because they have different amplitudes $A^{(+)}$ and $A^{(-)}$. This requirement together with the expression (\ref{omegaminusj}) of the frequencies implies that the amplitude vectors $A^{(+)}_{-}$ and $A^{(-)}_{-}$ at $x=-\infty$ should be related to the amplitude vectors $A^{(+)}_{+}$ and $A^{(-)}_{+}$ at $x=+\infty$
 by the simple relations
 \begin{equation}\label{amplirelminus}
 A^{(+)}_{+}= \Phi^{(+)} A^{(+)}_{-}\,\,,\,\,A^{(-)}_{+}= \exp(i\Theta^{(-)}) A^{(-)}_{-}\,\,,
 \end{equation}
 where $\Phi^{(+)}$ is a  constant $N^{(+)} \times N^{(+)}$ matrix in the $SL(N^{(+)})$ group defined by the condition 
\begin{equation}\label{groupminus}
\Phi^{(+)\dagger}S^{(+)} \Phi^{(+)} = S^{(+)}\,\,,
\end{equation}
and $\Theta^{(-)}$ is a constant real diagonal $N^{(-)}\times N^{(-)}$ matrix,  $\Theta^{(-)}=$ diag$\{\theta^{(-)}_1\,\cdots ,\,\theta^{(-)}_{N^{(-)}}\}$. This family of plane wave solutions (\ref{PWUexpminus}) and  (\ref{PWWexpminus}) allows for a complete definition of the class $\mathcal{D}$ of solutions of the equations 
(\ref{redboomeronminus}) in the first case.
\newline
In the second case the plane wave solutions are described by similar expressions, which read 
\begin{equation}\label{PWUexpplus}
\begin{array}{l}
U(x,t)= A^{(-)} A^{(+) \dagger}\exp[i(K^{(+)}x-\Omega^{(+)} t)]\,\,,
\end{array}
\end{equation}
while the $N^{(+)}\times N^{(+)}$ matrix $W^{(+)}(x,t)$ is found to be
\begin{equation}\label{PWWexpplus}
\begin{array}{l}
W^{(+)}(x,t)=i\exp[-i(K^{(+)}x-\Omega^{(+)} t)] B^{(+)}\exp[i(K^{(+)}x-\Omega^{(+)} t)]\,\,,  
\end{array}
\end{equation}
with the following specifications: $K^{(+)}$ is an arbitrary real constant diagonal  $N^{(+)}\times N^{(+)}$ matrix,
\begin{equation}\label{Kplus}
K^{(+)}=\text{diag}\{k^{(+)}_1,\cdots,k^{(+)}_{N^{(+)}}\}\,\,,
\end{equation}
with the assumption $k^{(+)}_j \neq k^{(+)}_n$ if $j\neq n$; the $N^{(+)}\times N^{(+)}$ matrix $\Omega^{(+)}$ is also a constant, real and diagonal matrix
\begin{equation}\label{Omegaplus}
\Omega^{(+)}=\text{diag}\{\omega^{(+)}_1,\cdots,\omega^{(+)}_{N^{(+)}}\}\,\,,
\end{equation}
whose entries take the expression
\begin{equation}\label{omegaplusj}
\omega^{(+)}_j=-c^{(+)}_j k^{(+)}_j - (A^{(-)}\,,\,S^{(-)} A^{(-)}) \sum_{n=1,n\neq j}^{N^{(+)}} S^{(+)}_n |A_n^{(+)}|^2 \left(\frac{c^{(+)}_j-c^{(+)}_n}{k^{(+)}_j-k^{(+)}_n}\right)\,\,;
\end{equation}
$A^{(+)}$ and $A^{(-)}$ are  again arbitrary constant vectors of dimension $N^{(+)}$ and, respectively,  $N^{(-)}$ and $B^{(+)}$ is the  $N^{(+)}\times N^{(+)}$ off--diagonal matrix whose  entries are
\begin{equation}\label{Bmatrixplus}
B^{(+)}_{jj}=0\,\,,\,\,B^{(+)}_{jn}= (A^{(-)}\,,\,S^{(-)} A^{(-)})S^{(+)}_j A_j^{(+)}A_n^{(+)*} \left(\frac{c^{(+)}_j-c^{(+)}_n}{k^{(+)}_j-k^{(+)}_n}\right)\,\,,\,\,j\neq n\,\,.
\end{equation}
Again these multi--component plane wave solutions are characterized by $N^{(+)}$ wave numbers $k^{(+)}_j$ and frequencies $\omega^{(+)}_j$, $j=1,\cdots,N^{(+)}$, and by $N^{(+)}$ amplitudes $A_m^{(+)}$ and $N^{(-)}$ amplitudes $A_j^{(-)}$, which are the components of the $N^{(+)}$--dimensional vector $A^{(+)}$ and, respectively, of the $N^{(-)}$--dimensional vector $A^{(-)}$.
Also in this case the definition of the class  $\mathcal{D}$ of solutions  for the special case   (\ref{redboomeronplus}) of the general equation (\ref{redboomeron}) is completed by  the  requirement  that the asymptotic plane wave solutions $U_{-PW}$ at $x=-\infty$ and $U_{+PW}$ at $x=+\infty$, see the definition (\ref{classD1}), have the same wave numbers $k^{(+)}_j$ and frequencies $\omega^{(+)}_j$ and may differ from each other only because they have different amplitudes $A^{(+)}$ and $A^{(-)}$. This requirement together with the expression (\ref{omegaplusj}) of the frequencies implies that the amplitude vectors $A^{(+)}_{-}$ and $A^{(-)}_{-}$ at $x=-\infty$ should be related to the amplitude vectors $A^{(+)}_{+}$ and $A^{(-)}_{+}$ at $x=+\infty$
 by the relations
  \begin{equation}\label{amplirel}
 A^{(-)}_{+}= \Phi^{(-)} A^{(-)}_{-}\,\,,\,\,A^{(+)}_{+}= \exp(i\Theta^{(+)}) A^{(+)}_{-}\,\,,
 \end{equation}
 where $\Phi^{(-)}$ is a  constant $N^{(-)} \times N^{(-)}$ matrix in the $SL(N^{(-)})$ group defined by the condition 
\begin{equation}\label{groupplus}
\Phi^{(-)\dagger}S^{(-)} \Phi^{(-)} = S^{(-)}\,\,,
\end{equation}
and $\Theta^{(+)}$ is a constant real diagonal $N^{(+)}\times N^{(+)}$ matrix,  $\Theta^{(+)}=$ diag$\{\theta^{(+)}_1\,\cdots ,\,\theta^{(+)}_{N^{(+)}}\}$. This family of plane wave solutions (\ref{PWUexpplus}) and  (\ref{PWWexpplus}) allows for a complete definition of the class $\mathcal{D}$ of solutions of the equations 
(\ref{redboomeronplus}) in the second case.

\noindent
The main ingredients of the algebraic construction of 
solutions $U(x,t)$, $W^{(\pm)}(x,t)$ are the \emph{seed} solutions $U^{(0)}(x,t)$, $W^{(0)(\pm)}(x,t)$ together with the
 corresponding solution $\psi ^{(0)}(x,t,k)$ of the Lax equations
(\ref{0Laxpair}) to which the
DDT applies (see (\ref{DDT})).  In its turn, the DDT
introduces additional parameters in the new solution, one being the pole
$\alpha$ in the complex $k$--plane, while other
parameters come from the residue matrix.
In this respect, a word of warning is appropriate. A generic choice of
these parameters does not necessarily yield a solution $U(x,t)$, $W^{(\pm)}(x,t)$ 
which is acceptable in the sense that i) it is bounded (i.e. it is 
singularity--free in the whole $(x,\,t)$ plane) and ii) it is localized (i.e. it
belongs to the same class, either $\mathcal{B}$ or $\mathcal{D}$, of
$U^{(0)}(x,\,t)$). Meeting these acceptability requirements is a property
of the new solution $U(x,t)$, $W^{(\pm)}(x,t)$ that has to be checked,
depending on all parameters which enter in the construction, including the sign matrices $S^{(+)}$ and
$S^{(-)}$. This analysis is quite simple in the
scalar case, while in the matrix case, which we are presently dealing
with, it requires some effort. Here we limit our attention to
the  case in which the seed solution $U^{(0)}(x,t)$, $W^{(0)(\pm)}(x,t)$ is
 the simplest solution of the functional class to which $U^{(0)}(x,t)$
belongs.  As a consequence, we obtain explicit expressions of the one--soliton solution.
Even so, the acceptability requirements mentioned above will be
investigated only a posteriori and in particularly interesting cases (see section \ref{3WRI}). 

\noindent
As for the pole $\alpha$ introduced by the DDT, see  (\ref{elementaryDDT}), we point out that its position, whether on the real axis or off the real axis, depends on the boundary values of $U(x,t)$ which characterize the functional classes $\mathcal{B}$ and $\mathcal{D}$. Indeed, in the construction of soliton solutions the pole $\alpha$ has a spectral meaning since it belongs to the discrete spectrum of the Lax operator $L(U)=-i\sigma\partial_x+i\sigma Q$, see the spectral equation (\ref{lax}) with (\ref{xLaxoperator}). In general, if $U(x,t)$ belongs to the class $\mathcal{B}$, then the discrete spectrum of the operator $L(U)$ cannot be on the real axis and only Darboux matrices $D$ with $\alpha\neq\alpha^{\ast}$ may give rise to physically meaningful solutions.  On the other hand, if $U(x,t)$ belongs instead to the class $\mathcal{D}$, then the correspending operator $L(U)$ may have gaps of the continuum spectrum on the real axis. In this case, the Darboux transformation, with $\alpha=\alpha^{\ast}$ lying in one of these gaps, yields meaningful solutions. The computation in this last case, which is precisely the one of dark solitons, is made difficult by the gap structure of the spectrum which gets more and more  complicate the higher the two integers $N^{(+)}$ and  $N^{(-)}$ are. 

%%%%%%%%%%%%%%%%sottosezione 3.1%%%%%%%%%%%%%%%%%%%%%%
\subsection{Bright solitons}\label{sec:bsoli}
Let us first consider solutions belonging to the  class $\mathcal{B}$. 
In this case the \emph{seed} solution of (\ref{redboomeron}) is $U^{(0)}(x,t)=0$ and $W^{(0)(\pm)}(x,t)=\mp F^{(\pm)}\,=$  constant matrix. The corresponding solution $\psi ^{(0)}(x,t,k)$ of the Lax equations (\ref{0Laxpair}) is
\begin{equation}\label{Bpsi0}
\psi ^{(0)}(x,t,k)=\left(\begin{array}{cc} \psi ^{(0)(+)}(x,t,k) & \mathbf{0} \\ \mathbf{0} & \psi ^{(0)(-)}(x,t,k)\end{array}\right)\,,
\end{equation}
where the blocks $\psi ^{(0)(\pm)}(x,t,k)$ are
\begin{equation}\label{Bpsi0pm}
\psi ^{(0)(\pm)}(x,t,k)=\exp[\pm ikx]\exp[(2ikC^{(\pm)}+F^{(\pm)})t]\,.
\end{equation}
We now follow the  DDT \emph{algorithm} of section \ref{sec:ddtalgo} by computing the vectors $z^{(\pm)}(x,t)$, see (\ref{Czsolution}), whose expression is
\begin{equation}\label{zBgood} 
z^{(\pm)}(x,t)=\exp(\pm i\alpha^{\ast} x)\,\Pi^{(\pm)}(t)\,z^{(\pm)}_0\,,
\end{equation}
where we have introduced the time--dependent matrices
\begin{equation}\label{PiB}
\Pi^{(\pm)}(t)=\exp[(2 i\alpha^{\ast} C^{(\pm)}+F^{(\pm)}) t]\,.
\end{equation}
In the present case, the DDT with a real pole,
$\alpha=\alpha^{\ast}$, has to be ruled out
since it leads to a solution $U(x,t)$, $W^{(\pm)}(x,t)$ which is not in the class $\mathcal{B}$ or $\mathcal{D}$. 
 Thus the pole $\alpha$ has to lie off the real axis, $\alpha \neq
\alpha^{\ast}$, and the DDT formulas (\ref{realBacklundcompl}) with
$U^{(0)}(x,t)=0$, $W^{(\pm)(0)}(x,t)=\mp F^{(\pm)}$  
yield the one--soliton solution
\begin{equation}\label{UBboomeron}
U(x,t)=-\frac{4b}{\Delta}\,\exp{(-2iax)}\, \Pi^{(-)}(t) B_0 \Pi^{(+)\dagger}(t) S^{(+)}\,\,,
\end{equation}
\begin{equation}\label{WpmBboomeron}
W^{(\pm)}(x,t)=\mp F^{(\pm)}-\frac{4b}{\Delta}\,\exp{(\pm2bx)}\, [C^{(\pm)},\Pi^{(\pm)}(t) B^{(\pm)}_0 \Pi^{(\pm)\dagger}(t) S^{(\pm)}]\,\,,
\end{equation}
where 
\begin{equation}\label{DeltaB}
\begin{array}{lll}
\Delta (x,t) & = & (z^{(+)},\,S^{(+)}z^{(+)})-(z^{(-)},\,S^{(-)}z^{(-)})=\\
& = & \exp{(2bx)} (z_0^{(+)}\,,\,\Pi^{(+)\dagger}(t)S^{(+)}\Pi^{(+)}(t) z^{(+)}_0 )\\
& - & \exp{(-2bx)}(z_0^{(-)}\,,\,\Pi^{(-)\dagger}(t)S^{(-)}\Pi^{(-)}(t) z^{(-)}_0 )\,\,.
\end{array}
\end{equation}
Here $\alpha=a+ib$, namely $a$ and $b$ are real parameters (the real and, respectively, imaginary part of the complex pole $\alpha$).  In these expressions we have also introduced for convenience the following constant matrices
\begin{equation}\label{B0s}
B_0=z_{0}^{(-)}z_{0}^{(+)\dagger}\,,\qquad B^{(\pm)}_0=z_{0}^{(\pm)}z_{0}^{(\pm)\dagger}\,.
\end{equation}
The expressions (\ref{UBboomeron}) and  (\ref{WpmBboomeron}) are an acceptable, i.e. bounded, solution if, and only if, the following inequality
\begin{equation}\label{denomcond}
(z_0^{(+)}\,,\,\Pi^{(+)\dagger}(t)S^{(+)}\Pi^{(+)}(t) z^{(+)}_0 )\,(z_0^{(-)}\,,\,\Pi^{(-)\dagger}(t)S^{(-)}\Pi^{(-)}(t) z^{(-)}_0 )\,<\,0
\end{equation}
holds at any time. For instance, if $S^{(+)}=1$ and $S^{(-)}=-1$, this condition is satisfied for any vector $z^{(\pm)}_0$.
 With this  condition, this multi--parameter solution $U(x,t)\,,\,W^{\pm}(x,t)$ is the one--soliton solution of the system (\ref{redboomeron}) which clearly belongs to the class $\mathcal{B}$. Indeed, while the expression (\ref{UBboomeron}) of $U(x,t)$ exponentially decays as $|x| \rightarrow \infty$, in this limit the matrix $W^{(\pm)}(x,t)$ goes to a matrix whose expression can be easily derived from the formulas (\ref {WpmBboomeron}) together with (\ref{DeltaB}), and which depends only on time. This special solution has been introduced in \cite{CD2004} for the system  (\ref{redboomeron}) with an additional dispersive term of Schr\"odinger--type. There it has been shown that these solutions exhibit a rich phenomenology as a consequence of the interplay of the  matrices $C^{(\pm)}$ and $F^{(\pm)}$. These soliton solutions may well feature processes such as decay, excitation, pair creation and annihilation,  and they are known, according to their behavior, as \emph{boomerons, trappons, simultons} (see also section 4 for additional references).
 
 %%%%%%%%%%%%%%%%sottosezione3.2%%%%%%%%%%%%%%%%%%%%%%
 \subsection{Dark solitons}\label{sec:dsoli}
 We now turn our attention to the one soliton solution which asymptotically behaves as a plane wave and is therefore  in the class $\mathcal{D}$. We limit the present construction of this solution to the evolution equations (\ref {redboomeronminus}), namely to the special case $C^{(+)}=W^{(+)}=0$. For this reason, and to simplify the notation, in this subsection we drop the superscript $(-)$ wherever reasonable, namely we set $C^{(-)}\equiv C\,,\,c^{(-)}_j\equiv c_j\,,\,K^{(-)}\equiv K\,,\,k_j^{(-)}\equiv k_j\,,\,\Omega^{(-)}\equiv \Omega\,,\,\omega_j^{(-)}\equiv \omega_j$ and $B^{(-)}\equiv B$.  The  construction of the one soliton solution of the equations (\ref {redboomeronplus}) in the other case $C^{(-)}=W^{(-)}=0$ is quite similar and is not reported. The solution obtains by applying the DDT to the plane wave background solution (\ref{PWUexpminus}) and   (\ref{PWWexpminus}) according to the formulae given in section 2, see (\ref{realBacklundcompl}) and (\ref{Rdenomin}). Therefore, we first  compute the corresponding expression of the solution $\psi ^{(0)}(x,t,k)$ of the Lax equations (\ref{0Laxpair}). To this aim we more conveniently write the seed matrices $Q^{(0)}(x,t)$ and $W^{(0)}(x,t)$ in the form  
 \begin{equation}\label{Qdnaked}
Q^{(0)}=G\mathcal{Q}G^{-1}\,,\quad W^{(0)}=G\mathcal{W}G^{-1}\,,
\end{equation}
where
\begin{equation}\label{mathcalG}
G=G(x,t)=\left(\begin{array}{cc} 1& \mathbf{0} \\ \mathbf{0} & \exp{[i(Kx-\Omega t)]}\end{array}\right)\,,
\end{equation}
\begin{equation}\label{M}
\mathcal{Q}=\left(\begin{array}{cc} \mathbf{0} & S^{(+)}A^{(+)}A^{(-){\dagger}}S^{(-)} \\ A^{(-)}A^{(+)\dagger} & \mathbf{0}\end{array}\right)\,,\quad \mathcal{W}=-i\left(\begin{array}{cc} \mathbf{0} & \mathbf{0} \\ \mathbf{0} & B\end{array}\right)\,.
\end{equation}  
This implies that the matrix 
\begin{equation}\label{phi}
\Phi (x,t,k)= G^{-1}(x,t)\psi ^{(0)}(x,t,k)
\end{equation}
satisfies the following pair of ODEs with constant coefficients
\begin{equation}\label{phi0x}
\Phi_x=i\,\mathcal{X}\,\Phi\,,\quad\Phi_t=-i\,\mathcal{T}\,\Phi\,.
\end{equation}
Here $\mathcal{X}$ and $\mathcal{T}$ are  $N\times N$ constant matrices ($N=N^{(-)}+N^{(+)}$) with the following block form
\begin{equation}\label{mathcalX}
\mathcal{X}(k)=\left(\begin{array}{cc}k & -iS^{(+)}A^{(+)}A^{(-){\dagger}}S^{(-)} \\ -iA^{(-)}A^{(+)\dagger} & -k-K\end{array}\right)\, \,, 
\end{equation}
\begin{equation}\label{mathcalT}
\mathcal{T}(k) = \left(\begin{array}{cc} \mathbf{0} & -iS^{(+)}A^{(+)}A^{(-){\dagger}}S^{(-)}C \\ -iCA^{(-)}A^{(+)\dagger} & -2kC+B-\Omega \end{array}\right)\,\,, 
\end{equation}
where all the quantities $A^{(\pm)}\,,\,B\,,\,K\,,\,\Omega$ are defined by the expressions (\ref{PWUexpminus}--\ref{Omegaminus}) of the naked solution $U^{(0)}\,,\,W^{(0)}$. Since the matrices $\mathcal{X}$ and $\mathcal{T}$ are constant and commute with each other, $[\,\mathcal{X}\,,\,\mathcal{T}\,]=0$, a fundamental solution of (\ref{phi0x}) is
\begin{equation}\label{phiexp}
\Phi(x,t,k)=\exp[i(x\,\mathcal{X}-t\,\mathcal{T})]\,\,,
\end{equation}
 and therefore the expression of $\psi ^{(0)}(x,t,k)$ reads
\begin{equation}\label{Dpsi0}
\psi^{(0)}(x,t,k)=G(x,t)\exp[i(x\,\mathcal{X}-t\,\mathcal{T})]\,\,.
\end{equation} 
In order to compute the one soliton solution $U\,,\,W$ by applying the DDT (\ref{realBacklundcompl}), the essential step is finding the explicit expression of the vectors $z^{(+)}$ and $z^{(-)}$. This amounts to choosing the constant vectors $z_0^{(+)}$ and $z_0^{(-)}$, as in (\ref{Czsolution}), in such a way that the resulting one soliton solution be in the class $\mathcal{D}$. To this aim we have to find the eigenvalues of the matrices $\mathcal{X}$ and $\mathcal{T}$ as well as their common eigenvectors. Let $\phi$ be an eigenvector of $\mathcal{X}$ and $\mathcal{T}$ and let $\chi$ and $\nu$ be the corresponding eigenvalues, namely
\begin{equation}\label{eigen}
\mathcal{X}\,\phi \,=\, \chi \, \phi \,,\quad \mathcal{T}\, \phi \,=\, \nu \,\phi \,\,.
\end{equation}
From the block expression (\ref{mathcalX}) of the matrix $\mathcal{X}$, the eigenvalue equation $\mathcal{X}\,\phi \,=\, \chi \, \phi$ may be rewritten as 
\begin{equation}\label{blockeigen}
\left \{ \begin{array}{l} (k-\chi ) \phi^{(+)} = i(A^{(-)}\,,\,S^{(-)} \phi^{(-)}) S^{(+)} A^{(+)} \\
(k+\chi +K) \phi ^{(-)} = -i(A^{(+)}\,,\,\phi^{(+)}) A^{(-)}
\end{array} \right .
\end{equation}
where, $ \phi ^{(\pm)}$ is the $N^{(\pm)}$--dimensional vector defined by our block notation
\begin{equation}\label{blockphi}
\phi = \left(\begin{array}{c}\phi^{(+)}  \\
\phi^{(-)} \end{array}\right)\,\,.
\end{equation}
It is plain that both the eigenvector $\phi=\phi(k)$, and therefore its blocks $\phi^{(+)}(k)$ and $\phi^{(-)}(k)$, and the eigenvalues $\chi=\chi(k)$ and $\nu=\nu(k)$ depend on the complex spectral variable $k$.     
Several consequences of these equations easily follow and are reported here below. 

\textbf{Proposition 1}~: if $N^{(+)}>1$ then there exist $N^{(+)}-1$ linearly independent eigenvectors such that $\phi^{(+)}$ is orthogonal to the vector $A^{(+)}$, i.e $(A^{(+)}\,,\,\phi^{(+)})=0$, and $\phi^{(-)} = 0$, while their corresponding eigenvalues, see (\ref{eigen}), are $\chi = k$ and $\nu= 0$ with multiplicity $N^{(+)}-1$.  
\newline
Because of this result, the task is that of finding the remaining $N^{(-)} + 1$ eigenvectors and eigenvalues.

\textbf{Proposition 2}~: if $\phi^{(+)}$ is not orthogonal to $A^{(+)}$, $(A^{(+)}\,,\,\phi^{(+)})\neq 0$, then the eigenvector takes the expression
\begin{equation}\label{eigenphi}
\phi = \left(\begin{array}{c}\phi^{(+)}  \\
\phi^{(-)} \end{array}\right)\,=\, \left(\begin{array}{c}i S^{(+)} A^{(+)}   \\
(A^{(+)}\,,\,S^{(+)} A^{(+)})(k+\chi +K)^{-1} A^{(-)} \end{array}\right)\,\,\,.
\end{equation}
Of course this expression is not explicit since it contains the unknown eigenvalue $\chi$. In its turn, this
 corresponding eigenvalue $\chi$ is one of the $N^{(-)} + 1$ solutions of the equation
\begin{equation}\label{eigenchi}
\chi = k- (A^{(+)}\,,\,S^{(+)} A^{(+)}) \sum_{j=i}^{N^{(-)}} S_{j}^{(-)}\frac{|A_{j}^{(-)}|^2}{\chi+k+k_{j}} 
\end{equation}
and the corresponding eigenvalue $\nu$, see (\ref{eigen}), is explicitly given as function of $\chi$ by the expression
\begin{equation}\label{eigennu}
\nu\,=\,- (A^{(+)}\,,\,S^{(+)} A^{(+)})  \sum_{j=i}^{N^{(-)}} S_{j}^{(-)}c_j \frac{|A_{j}^{(-)}|^2}{\chi+k+k_{j}}  \,\,. 
\end{equation}
For the sake of simplicity, from now on we assume that the $N^{(-)} + 1$ eigenvalues $\chi$ which solve the equation (\ref{eigenchi}) and the corresponding $N^{(-)} + 1$ eigenvalues $\nu$ given by (\ref{eigennu}), are all simple. Moreover we consider below only the case in which the spectral variable $k$ is real, $k=k^*$.

\textbf{Proposition 3}~: if $k=k^*$ and the eigenvalue $\chi$ is real, $\chi=\chi^*$, then also the eigenvalue $\nu$, corresponding to the same eigenvector, is  real, $\nu=\nu^*$.

\textbf{Proposition 4}~:  if $k=k^*$ and  $\chi$ and $\nu$ are eigenvalues corresponding to the same eigenvector, then also their complex conjugate $\chi^*$ and $\nu^*$ are eigenvalues.

This last result suggests the following notation: $N_c \geq 0$ is the number of eigenvalues $\chi=\chi_j\,, \,j=1,\cdots, N_c$, which lie in the upper half--plane, Im$\chi_j >0$. We then conclude that the $N^{(+)} + N^{(-)}$ eigenvalues of the matrix $\mathcal{X} (\mathcal{T})$ are divided as follows:  $N^{(+)} -1$ eigenvalues are explicitly given by Proposition 1, $2N_c$ eigenvalues come in complex conjugate pairs and the remaining $N^{(-)}+1-2N_c$ eigenvalues are real. In this respect, and for future reference, we observe that, because of the \emph{Hermitianity} relation
\begin{equation}\label{hermite}
\mathcal{X}^{\dagger}=\Sigma \mathcal{X} \Sigma\,\,,\quad
\mathcal{T}^{\dagger}=\Sigma \mathcal{T} \Sigma\,\,,
\end{equation}
where
\begin{equation}\label{Sigma}
\Sigma = \left(\begin{array}{lr} S^{(+)} & 0\\ 0 & -S^{(-)} \end{array} \right)\,\,,
\end{equation}
if $\Sigma = \pm \mathbf{1}$ it follows $N_c=0$.
Moreover, because of the properties (\ref{hermite}), the following orthogonality conditions hold true.

\textbf{Proposition 5}~:  if $k=k^*$, and if $\phi_j$ $(\hat{\phi}_j), j=1,\cdots,N_c$, denotes the eigenvector corresponding to the eigenvalues $\chi_j $ and $\nu_j$ ($\chi^*_j$ and $\nu^*_j$), then
\begin{equation}\label{orthog1}
(\phi_j\,,\,\Sigma \phi_n)= 0\,,\quad (\hat{\phi}_j\,,\,\Sigma \hat{\phi}_n)= 0\,,\quad j\,=1,\cdots,N_c\,,\quad n\,=1,\cdots,N_c\,,
\end{equation}
and
\begin{equation}\label{orthog2}
(\hat{\phi}_j\,,\,\Sigma \phi_n)= 0\,,\quad j \neq n\,\,.
\end{equation}
The only non vanishing scalar products are therefore $(\hat{\phi}_j\,,\,\Sigma \phi_j)\,,\, j=1,\cdots,N_c $. 

For future reference, we also observe the following.

\textbf{Proposition 6}~: if $k=k^*$ and $|k|$ is sufficiently large, then $N_c=0$ and the $N^{(-)} + 1$ eigenvalues $\chi(k)=\chi^*(k)$ are all real and have the following asymptotic value (see (\ref{eigenchi})): for one of them $\chi(k)=k +O(1/k)$, while for the others $N^{(-)}$ eigenvalues $\chi(k)=-k-k_j +O(1/k)$. 

By decreasing the value of $k$ two real eigenvalues may collide with each other and  create a pair of complex conjugate eigenvalues. These collisions of eigenvalues give rise to a complicate gap structure of the spectrum of the differential operator $L(U)= -i \sigma \partial_x +i \sigma Q$. Therefore, for a given real value of $k$, the generic vector solution of the Lax pair of equations (\ref{lax}) is a superposition of bounded oscillating solutions corresponding to real eigenvalues $\chi(k)$ and $\nu(k)$, and of exponentially unbounded solutions corresponding to  eigenvalues $\chi(k)$ and $\nu(k)$ which are instead complex. In order to construct a soliton solution in the class $\mathcal{D}$, only a linear combination of these last unbounded solutions can enter the DDT formulae of section 2.

\textbf{Proposition 7}~: if the pole $\alpha$ is real, $\alpha^*=\alpha$, then the vectors $z^{(+)}$ and  $z^{(-)}$ are given by the general equation (\ref{Czsolution}) with (\ref{Dpsi0}), namely 
 \begin{equation}\label{zvector}
 \left(\begin{array}{c}z^{(+)}(x,t)  \\
z^{(-)}(x,t) \end{array}\right)\,=\,G(x,t)\exp[i(x\,\mathcal{X}-t\,\mathcal{T})]\,\left(\begin{array}{c}z_0^{(+)}  \\
z_0^{(-)} \end{array}\right)\,\,,
\end{equation}
where $\mathcal{X}=\mathcal{X}(\alpha)$ and $\mathcal{T}=\mathcal{T}(\alpha)$, see (\ref{mathcalX}) and (\ref{mathcalT}). The constant vectors $z^{(+)}_0$ and $z^{(-)}_0$ are expressed by the linear superposition
 \begin{equation}\label{z0vector}
 \left(\begin{array}{c}z^{(+)}_0  \\
z^{(-)}_0 \end{array}\right)\,=\sum_{j=1}^{N_c} (\gamma_j \phi_j + \hat{\gamma}_j \hat{\phi}_j) \,\,,
\end{equation}
of the eigenvectors of  $\mathcal{X}(\alpha)$ and $\mathcal{T}(\alpha)$ corresponding to complex eigenvalues. Here $\gamma_j$ and  $\hat{\gamma}_j$ are complex coefficients which are constrained by equation (\ref{zcond}), together with Proposition 5, to satisfy the condition
\begin{equation}\label{gammacond1}
(z_0^{(+)},S^{(+)}z_0^{(+)})-(z_0^{(-)},S^{(-)}z_0^{(-)})=\sum_{j=1}^{N_c}[\gamma_j^*\hat{\gamma}_j (\phi_j\,,\,\Sigma \hat{\phi}_j)\,+\,c.c.\,]=0\,\,.
\end{equation}

As a consequence of this Proposition, the vectors $z^{(+)}(x,t)$ and $z^{(-)}(x,t)$ take the following expression
 \begin{equation}\label{zvector+}
 \left(\begin{array}{c}z^{(+)}(x,t)  \\
z^{(-)}(x,t) \end{array}\right)\,=\,G(x,t)\sum_{j=1}^{N_c} (\gamma_j e^{i\beta_j} \phi_j + \hat{\gamma}_j e^{i\beta^*_j} \hat{\phi}_j) \,\,,
\end{equation}
where we have introduced the complex phases
\begin{equation}\label{fasi}
\beta_j= \chi_j\, x - \nu_j\, t\,\,,\quad j=1,\cdots,N_c\,\,.
\end{equation}
Moreover, by taking into account the explicit block expression (\ref{eigenphi}), we finally obtain the expression of the vectors $ z^{(+)}$
\begin{equation}\label{z+}
z^{(+)}=i \sum_{j=1}^{N_c} (\gamma_j e^{i\beta_j}  + \hat{\gamma}_j e^{i\beta^*_j} )\,S^{(+)} A^{(+)}     \,\,,
\end{equation} 
and $z^{(-)}$
\begin{equation}\label{z-}
z^{(-)}= (A^{(+)}\,,\,S^{(+)} A^{(+)})e^{i(K x-\Omega t)} \sum_{j=1}^{N_c} [\gamma_j e^{i\beta_j} (\alpha+\chi_j +K)^{-1}  + \hat{\gamma}_j e^{i\beta^*_j} (\alpha+\chi_j^* +K)^{-1} ] A^{(-)} \,\,,
\end{equation} 
which enter the DDT formulae (\ref{realBacklundcompl}). From these formulae it is also clear that it remains to find the denominator function $\mathcal{E}(x,t)$ to complete the construction of the soliton solution. Since $\mathcal{E}_x= (1/2)g$, see (\ref{Rdenom}), and $g= 2(z^{(+)},S^{(+)}z^{(+)})$ because of the definition (\ref{fghmcomp}) and of the condition $(z^{(+)},S^{(+)}z^{(+)})-(z^{(-)},S^{(-)}z^{(-)})=0 \,,$ one has to first compute the function $g(x,t)$ whose expression reads
\begin{equation}\label{g}
\begin{array}{lll} 
g(x,t)&=&2(A^{(+)}\,,\,S^{(+)} A^{(+)})[ \sum_{j=1}^{N_c}\sum_{n=1}^{N_c} (\gamma_j \gamma_n^* e^{i(\beta_j -\beta_n^*)}  +\hat{\gamma}_j  \hat{\gamma}_n^* e^{i(\beta^*_j-\beta_n )})+\\ &+ &
  \sum_{n=1}^{N_c}\sum_{j\neq n=1}^{N_c}(\gamma_j\hat{ \gamma}_n^* e^{i(\beta_j -\beta_n)} + c.c.) + \sum_{j=1}^{N_c} (\gamma_j \hat{\gamma}_j^* + c.c.)] \,\,. \end{array}
\end{equation} 
Note that the RHS of this expression is the sum of exponentials and a term, the last one, which does not depend on $x$ and $t$. The function $m(x,t)$, see (\ref{fghmcomp}), is similarly expressed as a sum of exponentials and a constant term. However, because of the relations (see (\ref{Rdenomin})) $\mathcal{E}_x=g/2\,,\,\mathcal{E}_t=m$, only the constant term in the expression of $m(x,t)$ needs to be computed to obtain the explicit expression of $\mathcal{E}(x,t)$ which finally reads
 \begin{equation}\label{denom}
\begin{array}{lll}\mathcal{E}(x,t)&=&\mathcal{E}_0 -i (A^{(+)}\,,\,S^{(+)} A^{(+)})[ \sum_{j=1}^{N_c}\sum_{n=1}^{N_c} (\frac{\gamma_j \gamma_n^*}{\chi_j-\chi_n^*} e^{i(\beta_j -\beta_n^*)} +\frac{ \hat{\gamma}_j  \hat{\gamma}_n^*} {\chi_j^*-\chi_n}e^{i(\beta^*_j-\beta_n )})+\\ &+& 
  \sum_{n=1}^{N_c}\sum_{j\neq n=1}^{N_c}(\frac{\gamma_j\hat{ \gamma}_n^*}{\chi_j-\chi_n} e^{i(\beta_j -\beta_n)}  - c.c.)] + \frac{x}{2} (A^{(+)}\,,\,S^{(+)} A^{(+)})\sum_{j=1}^{N_c} (\gamma_j^* \hat{\gamma}_j + c.c.)\\ &-& t (A^{(+)}\,,\,S^{(+)} A^{(+)})^2 \sum_{j=1}^{N_c} [\gamma_j \hat{\gamma}^*_j 
  (A^{(-)}\,,\,S^{(-)}C(\alpha+\chi_j+K)^{-2} A^{(-)})+ c.c.] \,\,. \end{array}
\end{equation} 
This expression of $\mathcal{E}(x,t)$ greatly simplifies by requiring that it should not vanish for any $x$ at any fixed time $t$. In fact, if we order the eigenvalues $\chi_j=q_j+ip_j$ according to their imaginary part $p_j$ as $0<p_1<p_2<\cdots<p_{N_c}$, the denominator function $\mathcal{E}(x,t)$ exponentially diverges at both $x=\pm \infty$ with the leading term $-\frac{1}{2p_{N_c}}(A^{(+)}\,,\,S^{(+)} A^{(+)}) |\gamma_{N_c}|^2 \exp(-2p_{N_c}x) $ as $x\rightarrow -\infty$ and $\frac{1}{2p_{N_c}}(A^{(+)}\,,\,S^{(+)} A^{(+)}) | \hat{\gamma}_{N_c} |^2 \exp(2p_{N_c}x)$ as $x\rightarrow +\infty$. Since these two terms have opposite sign, one has to impose that either $\gamma_{N_c}=0$ or that $\hat{\gamma}_{N_c}=0$. Iterating this argument eventually leads to conclude that a necessary condition that the denominator $\mathcal{E}(x,t)$ never vanishes (alias that the soliton solution is not singular) is that either all coefficients $\gamma_j$ or all coefficients $\hat{\gamma}_j$ must vanish. Note that this constraint also implies  that  condition (\ref{gammacond1}) is automatically satisfied. We also observe that these two choices, namely $\hat{\gamma}_j=0$ and $\gamma_j=0$ formally obtain one from the other by changing the sign of all complex eigenvalues $ \chi_j$ and $\nu_j$, and therefore we consider only the first choice $\hat{\gamma}_j=0$. Thus we have the following.

\textbf{Proposition 8}~ :  the step--by--step method of dressing a plane wave solution to obtain a novel solution  amounts to inserting the expressions (see (\ref{z+}, \ref{z-}, \ref{denom})) 
\begin{subequations}\label{zdenfin}
\begin{equation}\label{z+fin}
z^{(+)}=i \sum_{j=1}^{N_c} \gamma_j e^{i\beta_j} \,S^{(+)} A^{(+)}   \,\,,
\end{equation}
\begin{equation}\label{z-fin}
z^{(-)}= (A^{(+)}\,,\,S^{(+)} A^{(+)})e^{i(K x-\Omega t)} \sum_{j=1}^{N_c} \gamma_j e^{i\beta_j} (\alpha+\chi_j +K)^{-1} \, A^{(-)} \,\,,
\end{equation} 
 \begin{equation}\label{denfin}
\mathcal{E}(x,t)=\mathcal{E}_0 -i (A^{(+)}\,,\,S^{(+)} A^{(+)})\, \sum_{j=1}^{N_c}\sum_{n=1}^{N_c} \frac{\gamma_j \gamma_n^*}{\chi_j-\chi_n^*} e^{i(\beta_j -\beta_n^*)} 
\end{equation} 
\end{subequations}
in the general formulae (\ref{realBacklundcompl}) with 
\begin{subequations}\label{UWnaked}
\begin{equation}\label{Unaked}
U^{(0)}(x,t)=  e^{i(Kx-\Omega t)}  A^{(-)} A^{(+)\dagger}\,\,\,,
\end{equation}
\begin{equation}\label{Wnaked}
W^{(0)(+)}(x,t)=0\,\,\,,\,\,\, W^{(0)(-)}(x,t)= -i e^{i(Kx-\Omega t)} Be^{-i(Kx-\Omega t)} \,\,\,,
\end{equation} 
\end{subequations}
and the constant matrix $B=B^{(-)}$ as given by (\ref{Bmatrixminus}).

We close this section by showing that this novel solution we have constructed is, if locally bounded, in the class $\mathcal{D}$. To this aim we compute the asymptotic behavior of the matrices $U(x,t)$ and $W^{(-)}(x,t)$ for fixed $t$ and for $x\rightarrow \pm \infty$. This computation is straight  and it yields the following expressions
\begin{subequations}\label{UWasymp}
\begin{equation}\label{Uasymp}
U(x,t)\rightarrow  e^{i(Kx-\Omega t)}  A_{\pm}^{(-)} A^{(+)\dagger}\,,\,\quad x \rightarrow \pm \infty\,\,,
\end{equation}
\begin{equation}\label{Wasymp}
W^{(-)}(x,t) \rightarrow -i e^{i(Kx-\Omega t)} B_{\pm} e^{-i(Kx-\Omega t)} \,,\, \quad                                                                     x \rightarrow \pm \infty\,\,,
\end{equation} 
\end{subequations}
where the asymptotic constant vectors $A_{\pm}^{(-)}$ and matrices $ B_{\pm}$ are given by
\begin{equation}\label{ABasymp}
A_+^{(-)} = A^{(-)}\,,\quad A_-^{(-)} = \Gamma A^{(-)}\,,\quad B_{+}=B\,,\,B_{-}=\Gamma B \Gamma^{\dagger}\,.
\end{equation}
Here we have introduced the diagonal unitary matrix $\Gamma$
\begin{equation}\label{Gamma}
\Gamma= (\alpha + \chi_{N_c}^* + K)\,(\alpha + \chi_{N_c} + K)^{-1}= \mathrm{diag} \{e^{i\theta_1},\cdots,e^{i\theta_{N_c}} \}\,\,,
\end{equation}
where  the $N_c$ phases $\theta_j$ are defined as
\begin{equation}\label{theta}
\tan(\theta_j)= - \frac{p_{N_c}}{\alpha + q_{N_c} +k_j}\,\,,\quad j=1,\cdots,N_c\,\,.
\end{equation}
 Therefore, as displayed by our notation, the $U$ and $W$ parts of the solution are asymptotically  plane waves of the expected form $\exp[i(Kx-\Omega t)] A^{(-)}_{\pm} A^{(+)\dagger}$ and, respectively, $-i e^{i(Kx-\Omega t)} B_{\pm} e^{-i(Kx-\Omega t)}$  as $x \rightarrow \pm \infty$ where the asymptotic constant vectors $A^{(-)}_{+}$ and $A^{(-)}_{-}$ and matrices $B_{+}$ and $B_{-}$ are related to each other by the unitary transformation (\ref{Gamma}), namely $A^{(-)}_{-}=\Gamma A^{(-)}_{+}$, and, similarly, $B_{-}=\Gamma B_{+} \Gamma^{\dagger}$, where $\Gamma \Gamma^{\dagger}=1$.
 In the following section we apply the DDT one soliton solution construction to the simplest case $N^{(+)}=1$, $N^{(-)}=2$, in both classes $\mathcal{B}$ and $\mathcal{D}$. The choice $N^{(+)}=1$, $N^{(-)}=2$ leads to the three wave resonant interaction model. 
  
%%%%%%%%%%%%%%%%%%sezione4%%%%%%%%%%%%%%%%%%%%%%
\section{The 3WRI model}\label{3WRI}
The simplest and yet celebrated  system of partial differential equations which serves as particular example of the results presented in the previous sections is the one describing the resonant interaction of three waves in a medium with quadratic nonlinearity. In our notation of section 2 this system is obtained by choosing, for instance, $N^{(+)}=1\,,\,N^{(-)}=2\,,\,C^{(+)}=0\,,\,C^{(-)}=C=C^{\ast}=\,$diag$\{c_1,c_2\},c_1\neq c_2\,,\,S^{(+)}=1\,,\,S^{(-)}=S=\,$diag$\{s_1,s_2\}$, and by asking that the solution be of the form
\begin{equation}\label{3wred}
U(x,t)= \left ( \begin{array}{c} u_1(x,t) \\ u_2(x,t) \end{array} \right )\,,\,\,W^{(+)}(x,t)=0\,,\,\,W^{(-)}(x,t)=\left ( \begin{array}{cc} 0&-s_2w^{\ast}(x,t)\\s_1 w(x,t)&0  \end{array} \right )\,.
\end{equation}
In this case the general matrix PDEs (\ref{redboomeron}) reduce to the following system of three coupled first order equations (3WRI equations)
\begin{equation}\label{3wri}
u_{1t}+c_1u_{1x}=-s_2w^*u_2\,,\,\,\,u_{2t}+c_2u_{2x}=s_1wu_1\,,\,\,\,w_x=(c_2-c_1)u_1^*u_2\,,
\end{equation}
whose integrability was first proved in \cite{ZM}.
It should be pointed out that this system, being first order, is covariant with respect to arbitrary linear transformations of the coordinate plane $(x\,,\,t)$ in itself. This implies that one can give it the form, and attach to the independent variables $x\,,\,t$ the physical meaning, which is appropriate to the particular application of interest. Therefore we discuss the system (\ref{3wri}) with the understanding that all formulae we display here can be easily translated into corresponding results for any other form of the 3WRI equations which is covariantly related to it. 
\newline
Let us  consider first the one soliton solution in the class $\mathcal{B}$. This solution is constructed by mere application of the DDT formulae corresponding to a complex pole $\alpha=a+ib$, namely (\ref{UBboomeron}), (\ref{WpmBboomeron}) and  (\ref{DeltaB}) which in this case read
\begin{equation}\label{UB3wri}
U(x,t)=-\frac{4bz^{(+)*}_0}{\Delta}\,\exp{(-2iax)}\, \Pi(t) z^{(-)}_0\,\,,
\end{equation}
\begin{equation}\label{WB3wri}
w(x,t)= f-\frac{4b(c_2-c_1)}{\Delta}\,\exp{(-2bx)}\, [\Pi(t) z^{(-)}_0z^{(-)\dagger}_0 \Pi^{\dagger}(t)]_{21}\,\,,
\end{equation}
where 
\begin{equation}\label{DeltaB3wri}
\Delta (x,t)  =  \exp{(2bx)} |z_0^{(+)}|^2
  -  \exp{(-2bx)}(z_0^{(-)}\,,\,\Pi^{\dagger}(t)S\Pi(t) z^{(-)}_0 )\,\,.
\end{equation}
The constant complex parameters which characterize this solution are: $z_0^{(+)}$, the two-dimensional vector $z^{(-)}_0$ and $f$ which comes from the naked solution (see(\ref{WpmBboomeron}))
\begin{equation}\label{wnaked}
F^{(-)}\equiv F=\left ( \begin{array}{cc} 0&-s_2f^*\\s_1 f&0  \end{array} \right )\,\,. 
\end{equation}
Moreover the $2\times 2$ time--dependent  matrix $\Pi(t)$ is (see (\ref{PiB}))
\begin{equation}\label{Pi3wri}
\Pi(t)=\exp[(2i\alpha^* C+F)t]
\end{equation}
and the notation $[M]_{21}$ in the right hand side of (\ref{WB3wri})  indicates the matrix element $M_{21}$ of the matrix $M$. From the expression (\ref{DeltaB3wri}) of the denominator function $\Delta (x,t)$ it follows that this soliton solution is singular, due to the zeros of this denominator, if the signs $s_1$ and $s_2$ are $s_1=s_2=1$,  $s_1=-s_2=1$ and $s_1=-s_2=-1$ and it is instead regular, i.e. bounded in the entire $(x,t)$ plane, only if $s_1=s_2=-1$. In the particular case in which the parameter $f$ vanishes, $f=0$, this solution has been first introduced in \cite{ZM} and fully investigated in \cite{Kaup}--\cite{DL}. It describes the interaction of three bright-type pulses, and its experimental observation in optics has been recently reported \cite{bcacdwbd}. The case corresponding to $f\neq 0$ has been discovered and investigated later in \cite{CD2006} where the soliton behavior has been shown to be of boomeronic type \cite{DL3WRI}. These soliton solutions describe the interaction of two bright pulses $u_1(x,t)$ and $u_2(x,t)$ and a third pulse, $w(x,t)$, which is instead of dark type as it does not vanish when $x\rightarrow \pm \infty$. Corresponding to various choices of the linear group velocities $c_1\,,\,c_2$, this solution features a rich phenomenology which has been recently shown to be of applicative interest in optical processes in quadratically nonlinear media \cite{dcbw}--\cite{cbdw2008}. The properties of this soliton solution do not need to be reported  here since they have been detailed in the literature quoted above. 
\newline
Let us turn our attention to the soliton solution in the class $\mathcal{D}$. We consider only those solutions which are obtained by applying the DDT technique with a real pole $\alpha=\alpha^*$, see (\ref{elementaryDDT}), and this amounts to dressing a plane wave solution. Thus we apply to the 3WRI system (\ref{3wri}) the general formulae displayed in subsection 3.2 as follows. By setting $A^{(+)}=1$ and $A^{(-)}=A$, this being a 2--dimensional constant complex vector, in (\ref{Unaked}), (\ref{Wnaked}) and (\ref{Bmatrixminus}), the naked plane wave solution takes the expression
\begin{equation}\label{pw}
u^{(0)}_j= e^{[i(k_j x-\omega_j t)]} A_j\,,\,\,j=1,2\,,\quad  w^{(0)}=-i\left(\frac{c_2-c_1}{k_2-k_1}\right)A_2 A_1^*e^{\{i[(k_2-k_1)x-(\omega_2-\omega_1)t]\}}\,\,,
\end{equation}
where $k_1$ and $k_2$ are arbitrary real wave numbers and the corresponding frequencies $\omega_1, \omega_2$ are
\begin{equation}\label{omegas}
\omega_1=c_1 k_1+s_2|A_2|^2\left(\frac{c_2-c_1}{k_2-k_1}\right)\,\,\,,\,\,\,\omega_2=c_2 k_2+s_1|A_1|^2\left(\frac{c_2-c_1}{k_2-k_1}\right)\,\,.
\end{equation}
The continuous spectrum of the spectral problem $\psi^{(0)}_x=X^{(0)}\psi^{(0)}$, see (\ref{0Laxpair}), which corresponds to this plane wave may either coincides with the entire real $k$--axis, or may be the real axis with one or two gaps (see below). A gap $\mathcal{I}_g=\{k: k_{-}\leq k\leq k_{+}\}$ of the continuous spectrum is defined as an interval of the real axis such that, for $k\in \mathcal{I}_g$, the $3\times 3$ matrix $\mathcal{X}(k)$, see (\ref{mathcalX}),
\begin{equation}\label{3wrimathcalX}
\mathcal{X}(k)=\left(\begin{array}{ccc}k & -is_1A_1^* & -is_2A_2^* \\ -iA_1 & -k-k_1 & 0\\ -iA_2 & 0 & -k-k_2\end{array}\right)\, \,
\end{equation}
has one real eigenvalue and two complex (conjugate) eigenvalues, in contrast with the continuous spectrum where the three eigenvalues are all real. The very existence of a gap depends on the given data $k_1$,  $k_2$, $A_1$,  $A_2$, $s_1$, $s_2$. For instance if $s_1=s_2=-1$ the matrix (\ref{3wrimathcalX}) is Hermitian and therefore no gap may occur. In the present case the eigenvalues, which solve the equation (\ref{eigenchi}), are the roots of the third degree polynomial 
\begin{equation}\label{poly}
\mathcal{P}(\chi)= (\chi-k)^3+d_2 (\chi-k)^2 +d_1 (\chi-k) + d_0
\end{equation}
where
\begin{equation}\label{coeffi}
%\left \{
\begin{array}{lll} d_2(k)&=& 4k + k_1 +k_2\,\,,\\d_1(k)&=& 4k^2 + 2k(k_1+k_2) + k_1k_2 +s_1|A_1|^2 +s_2|A_2|^2\,\,,\\ d_0(k)& =& 2k (s_1|A_1|^2 +s_2|A_2|^2) + s_1k_2 |A_1|^2 +s_2k_1|A_2|^2\,\,.\end{array} 
%\right .
\end{equation}
and the eigenvalues may be explicitly given by  Cardano formulae. However we skip such details  and we limit our present discussion to the observation that the spectrum can have at most \emph{two} gaps. This result follows from the expression DIS$(k)$ of the discriminant of the equation $\mathcal{P}(\chi)=0$ which comes to be a fourth degree polynomial of the spectral variable $k$, namely
\begin{equation}\label{dis}
\mathrm{DIS}(k)= -\frac{4}{27}(k_1-k_2)^2 k^4+D_3 k^3 +D_2 k^2 + D_1 k + D_0\,\,\,,
\end{equation}
where the real coefficients D$_n$ depend only on the given data $k_1$,  $k_2$, $A_1$,  $A_2$, $s_1$, $s_2$. Indeed, since the three eigenvalues, namely the roots of $\mathcal{P}(\chi)$, are all real if the discriminant is negative, $\mathrm{DIS}(k)<0$, while two of them are complex conjugate if the discriminant is positive, $\mathrm{DIS}(k)>0$, the three eigenvalues are certainly all real for large enough $|k|$ because the coefficient of the highest power of  $\mathrm{DIS}(k)$ is negative (see also Proposition 6). Therefore, if the four roots of  $\mathrm{DIS}(k)$ are all complex (i.e. with a non vanishing imaginary part) then there are no gaps in the spectrum. If instead only two roots are real, 
$\mathrm{DIS}(k_-)=\mathrm{DIS}(k_+)=0$, $k_-\,<\,k_+$, the spectrum has one gap $k_-\,\leq \,k\,\leq k_+$ with these two roots $k_{\pm}$ being the endpoints of the gap interval. Finally, if all roots of $\mathrm{DIS}(k)$ are real, $\mathrm{DIS}(k^{(1)}_-)=\mathrm{DIS}(k^{(1)}_+)=\mathrm{DIS}(k^{(2)}_-)=\mathrm{DIS}(k^{(2)}_+)=0$, $k^{(1)}_-\,<\,k^{(1)}_+\,<\,k^{(2)}_-\,<\,k^{(2)}_+$, the spectrum has two gaps, $k^{(1)}_-\,\leq \,k\,\leq k^{(1)}_+$ and  $k^{(2)}_-\,\leq \,k\,\leq k^{(2)}_+$ (see Fig. 1).

\begin{figure}[h!]\label{fig0}
\centering
\fbox{\includegraphics[totalheight=3.5cm]{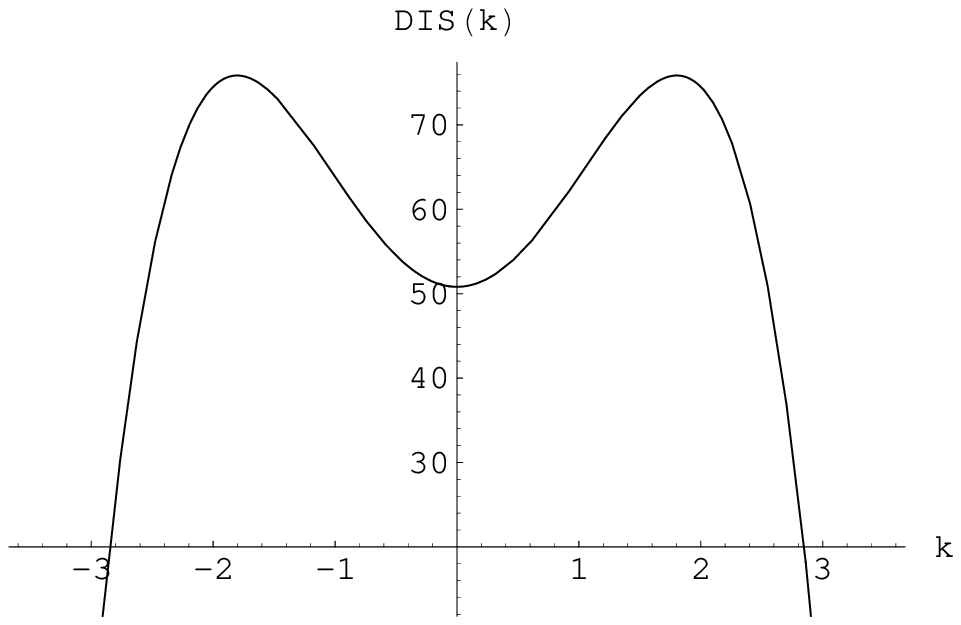}}
\hspace{0.2cm}
\fbox{\includegraphics[totalheight=3.5cm]{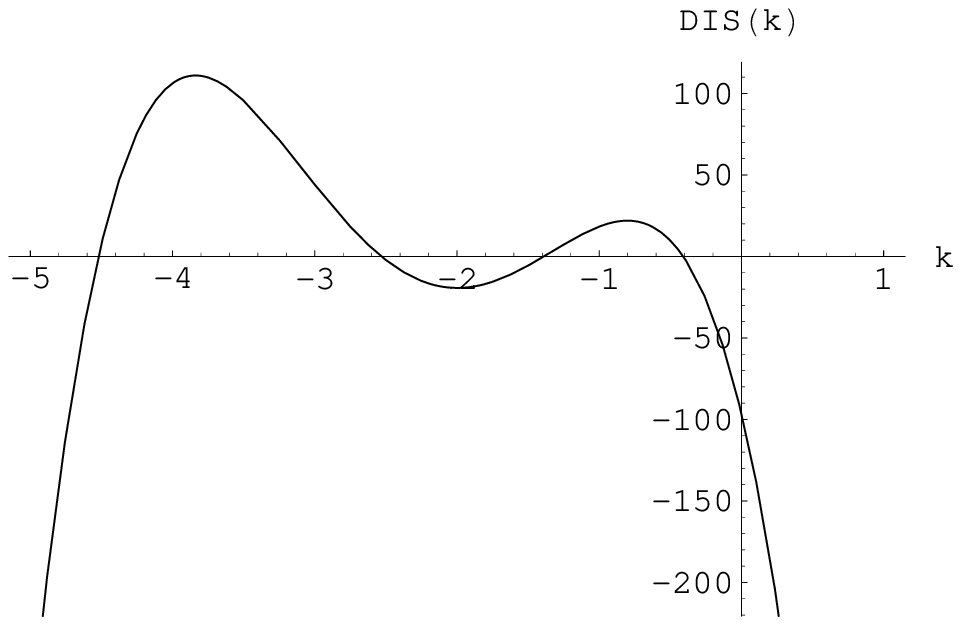}}
\caption{One or two gaps of the spectrum. In both pictures $s_1=s_2=1$. On the left: $k_1=-k_2=1$, $A_1=A_2=2$ and the gap is given by $k_{-}\leq k\leq k_{+}$, where $k_{\pm}=\pm\frac{1}{2}\sqrt{13+16\sqrt{2}}$. On the right:  $k_1=7$, $k_2=2$, $A_1=1$, $A_2=1/2$ and the two gaps are characterised by $k^{(1)}_{-}=-4.5$, $k^{(1)}_{+}=-2.6$ and $k^{(2)}_{-}=-1.5$, $k^{(2)}_{+}=-0.5$.
}
\end{figure}

In order to proceed further with our construction of the soliton solution, we assume the given data $k_1$,  $k_2$, $A_1$,  $A_2$, $s_1$, $s_2$ are such that the continuous spectrum has at least one  gap $\mathcal{I}_g$ and that the real pole $\alpha$ of the DDT lies inside this gap, $\alpha \in \mathcal{I}_g$.

We specialize further the expressions (\ref{zdenfin}) to the present case, namely
\begin{equation}\label{3wriz}
z^{(+)}=i\gamma e^{i\beta},\,\,\,z_j^{(-)}=\frac{\gamma e^{i(\beta +k_j x-\omega_j t)}}{\alpha +k_j +q+ip} A_j\,,\,\,\,j=1,2 \,,\quad \mathcal{E}=\mathcal{E}_0- \frac{|\gamma|^2}{2p}e^{-2\mathrm{Im}\beta}\,\,,
\end{equation}
 with the following specifications: $\chi=q+ip$ is the complex eigenvalue of the matrix $\mathcal{X}(\alpha)$, see (\ref{3wrimathcalX}), which lies in the upper half of the complex plane (i.e. $p>0$), $\beta=\chi x - \nu t$ and $\nu$ is the corresponding $\mathcal{T}(\alpha)$ eigenvalue, see (\ref{eigen}), given by (\ref{eigennu}), namely
 \begin{equation}\label{3wrinu}
 \nu=-\frac{s_1 c_1}{\alpha +q +k_1+ip} |A_1|^2 -\frac{s_2 c_2}{\alpha +q +k_2+ip} |A_2|^2\,\,.
 \end{equation}
 Inserting these expressions in the general DDT formula (\ref{realBacklundcompl}), and using the notation $\mathcal{E}_0= -|\gamma|^2 \exp(-2\xi_0)/(2p)$ and the new variable
 \begin{equation}\label{xicoord}
 \xi= \mathrm{Im}\,\beta - \xi_0=p(x-Vt)-\xi_0
 \end{equation}
 finally yields the one soliton expression
 \begin{subequations}\label{DDD}
\begin{equation}\label{uDDD}
u_j= A_j\, e^{i(k_j x-\omega_j t)} \left(\frac{e^{\xi}+e^{2i\theta_j}\, e^{-\xi}}{e^{\xi}+ e^{-\xi}}\right),\quad j=1,2\,\,,
\end{equation}
 \begin{equation}\label{wDDD}
 w=-i\left(\frac{c_2-c_1}{k_2-k_1}\right)A_2 A_1^{\ast} e^{i[(k_2-k_1)x-(\omega_2-\omega_1)t]} \left(\frac{e^{\xi}+e^{2i(\theta_2-\theta_1)}\, e^{-\xi}}{e^{\xi}+ e^{-\xi}}\right)\,\,.
\end{equation}
\end{subequations}
 This solution is a \emph{dark simulton} as it describes three pulses travelling on a non vanishing asymptotic plateau $(|u_j| \approx |A_j|\,,\,|w|\approx |(c_2-c_1)/(k_2-k_1)|A_1||A_2|)$ and locked toghether to travel at the same velocity $V$, see (\ref{xicoord}). This common velocity is obtained via the definition of the variable $\xi$, which implies Im$\,\nu$=pV, and  it takes the expression (see (\ref{3wrinu}))
 \begin{equation}\label{velo}
 V=\frac{s_1 c_1|A_1|^2}{(\alpha +q +k_1)^2+p^2} + \frac{s_2 c_2|A_2|^2}{(\alpha +q +k_2)^2+p^2} \,\,.
 \end{equation} 
This expression is not explicit since it gives $V$ in terms of the eigenvalue $\chi=q+ip$;
 we note however that it implies the following:
 
\textbf{Proposition 9}~: if we assume, for definiteness, the ordering $c_1<c_2$, then, according to the signs $s_1$ and $s_2$, we find that
 \begin{equation}\label{++}
 c_1\,<\,V_-\,\leq V\leq V_+\,<\,c_2\,\,,\,\,\mathrm{if}\,\,s_1=s_2=1\,\,\,,
 \end{equation}
 \begin{equation}\label{+-}
 V=c_1-(c_2-c_1)\frac{|A_2|^2}{(\alpha +q +k_2)^2+p^2}\,\,,\,\,
 -\infty\,<\,V_-\,\leq V\leq V_+\,<c_1\,\,,\,\,\mathrm{if}\,\,s_1=1\,,\,s_2=-1\,\,\,,
 \end{equation}
 \begin{equation}\label{-+}
 V=c_2+(c_2-c_1)\frac{|A_1|^2}{(\alpha +q +k_1)^2+p^2}\,\,,\,\,
 c_2\,<\,V_-\,\leq V\leq V_+\,<+\infty\,\,,\,\,\mathrm{if}\,\,s_1=-1\,,\,s_2=1\,\,\,.
\end{equation}

\begin{figure}[h!]\label{fig1}
\centering
\fbox{\includegraphics[totalheight=4cm]{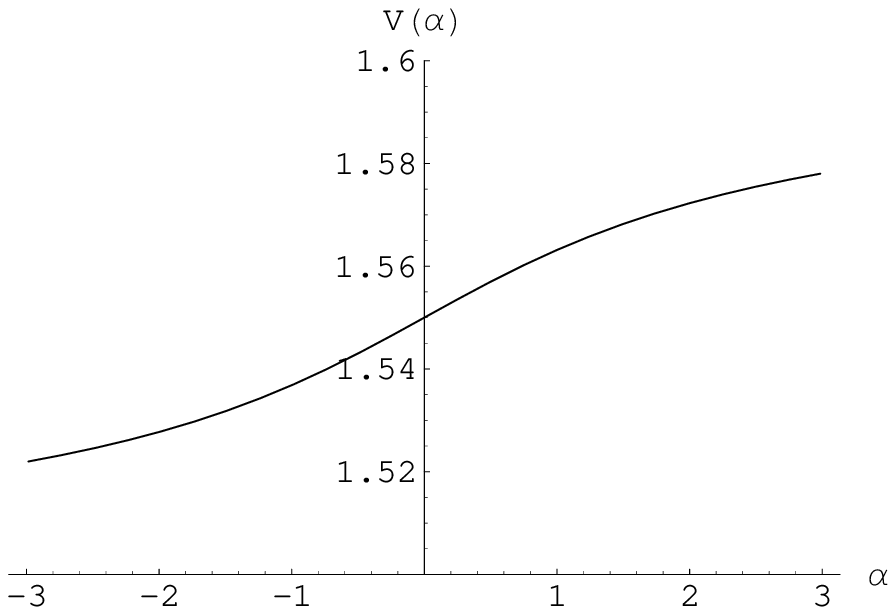}}
\hspace{0.2cm}
\fbox{\includegraphics[totalheight=4cm]{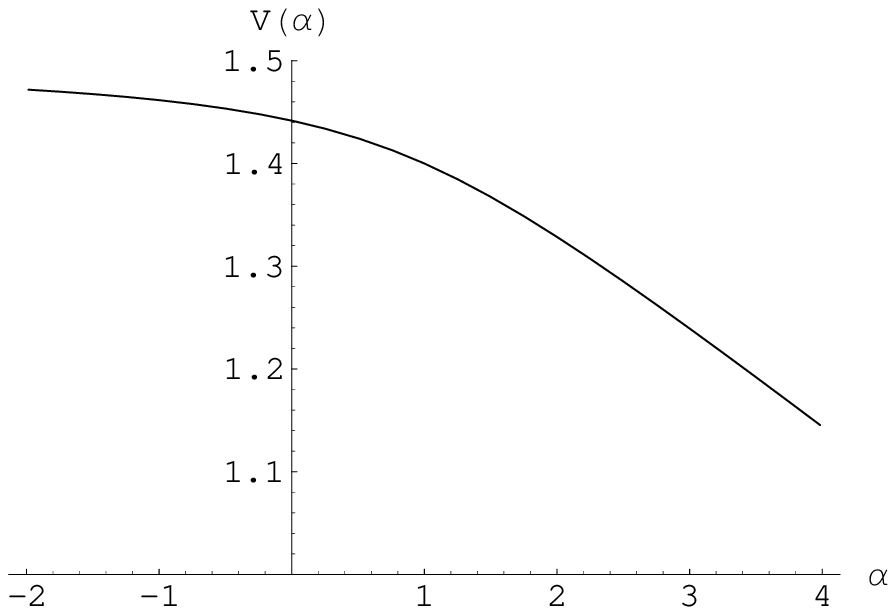}}
\caption{$k_1=-k_2=1$, $A_1=A_2=2$, $c_1=1.5$, $c_2=1.6$. On the left: $s_1=s_2=1$ and $ c_1\,<\,V_-\,\leq V\leq V_+\,<\,c_2$.  On the right: $s_1=-s_2=1$ and $ -\infty\,<\,V_-\,\leq V\leq V_+\,<c_1$.}
\end{figure}

The results of proposition 9 are a strait consequence of the relation
 \begin{equation}\label{iden}
 1=\frac{s_1 |A_1|^2}{(\alpha +q +k_1)^2+p^2} + \frac{s_2 |A_2|^2}{(\alpha +q +k_2)^2+p^2} \,,
 \end{equation}
 which follows from the imaginary part of the equation
   \begin{equation}\label{3wrichi}
 q+ip=\alpha-\frac{s_1 }{\alpha +q +k_1+ip} |A_1|^2 -\frac{s_2 }{\alpha +q +k_2+ip} |A_2|^2\,,
 \end{equation}
 this being just the present case form of  (\ref{eigenchi}). 
 We also note that the relation (\ref{iden}) shows again that this soliton solution does not exists if $s_1=s_2=-1$.
 
 Figure 2 shows an example in which only one gap is present.
 If two gaps $\mathcal{I}^{(1)}_g$ and $\mathcal{I}^{(2)}_g$ exist, where $\mathcal{I}^{(i)}_{g}$ is characterised by $k^{(i)}_{-}\leq k\leq k^{(i)}_{+}\,\,,\,\,i=1\,,\,2$, the results are similar, e.g. in the case of $s_1=s_2=1$ the velocity of the simulton $V$ lies in an interval when $\alpha$ varies in the gaps; however the function $V(\alpha)$ (together with its range of values) is different in the two gaps (see Figure 3).

\begin{figure}[h!]\label{fig2}
\centering
\fbox{\includegraphics[totalheight=4cm]{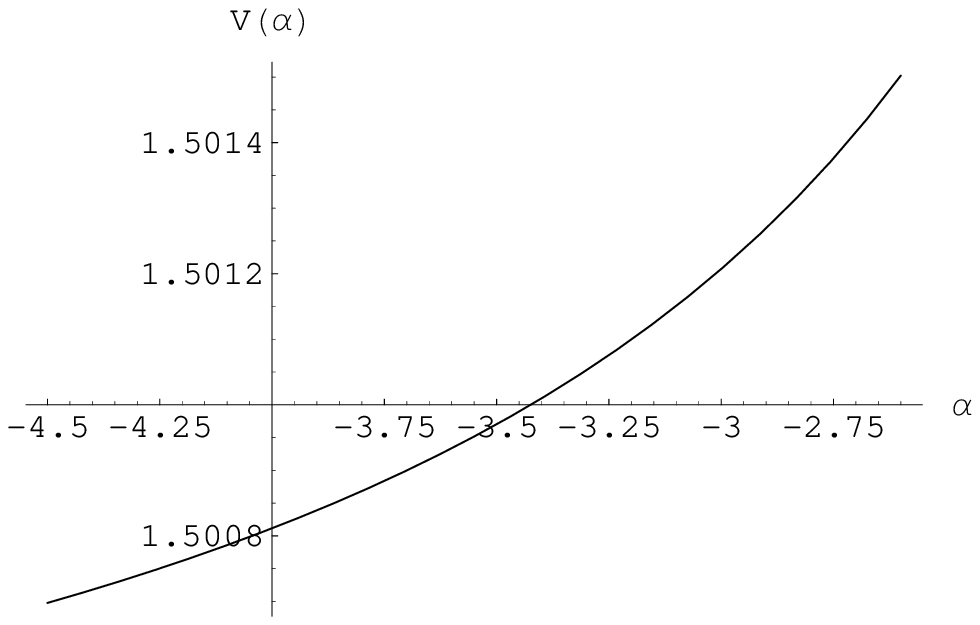}}
\hspace{0.2cm}
\fbox{\includegraphics[totalheight=4cm]{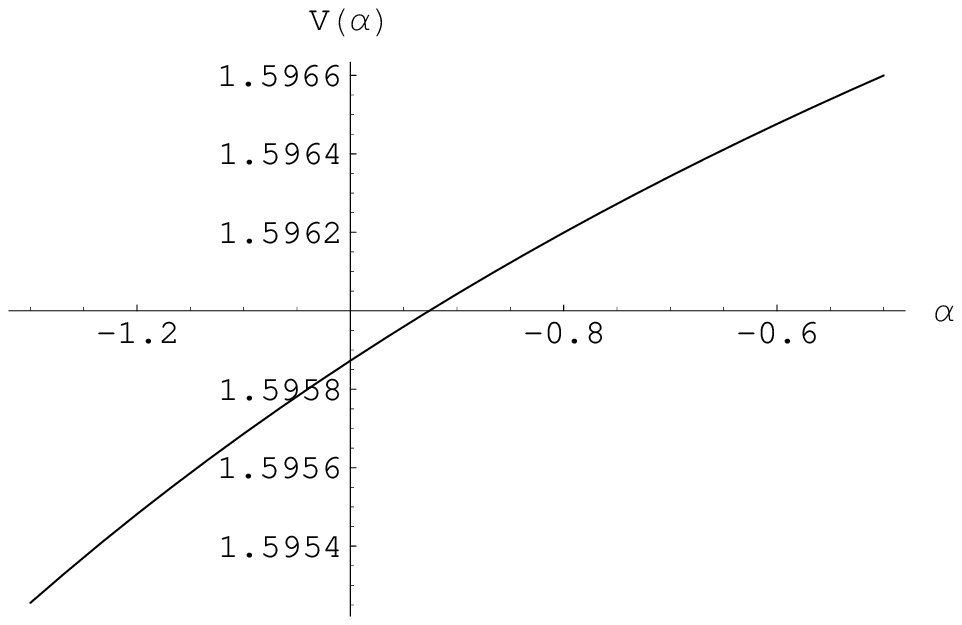}}
\caption{$k_1=7$, $k_2=2$, $A_1=1$, $A_2=1/2$, $c_1=1.5$, $c_2=1.6$; $s_1=s_2=1$. On the left: $k^{(1)}_{-}=-4.5$, $k^{(1)}_{+}=-2.6$; on the right: $k^{(2)}_{-}=-1.5$, $k^{(2)}_{+}=-0.5$. In both gaps $c_1\,<\,V_-\,\leq V\leq V_+\,<\,c_2 $.}
\end{figure}

Finally, it should be pointed out that this solution, as explicitly shown by (\ref{DDD}), describes the nonlinear interaction of three plane waves whose amplitudes asymptotically, namely at $t=\pm \infty$, acquire a  phase--shift. The expression of these phase--shifts is explicitly given by the general formula (\ref{theta}) which now reads
 \begin{equation}\label{3writheta}
\tan(\theta_j)= - \frac{p}{\alpha + q +k_j}\,\,,\,\,j=1,2\,\,.
\end{equation}
As an instance, the dependence of the phase--shifts $\theta_1$ and $\theta_2$ on the parameter $\alpha$ inside the gap, $k_-\,<\,\alpha\,<\,k_+$, is plotted in Figure 4.

\begin{figure}[h!]\label{fig3}
\centering
\fbox{\includegraphics[totalheight=6cm]{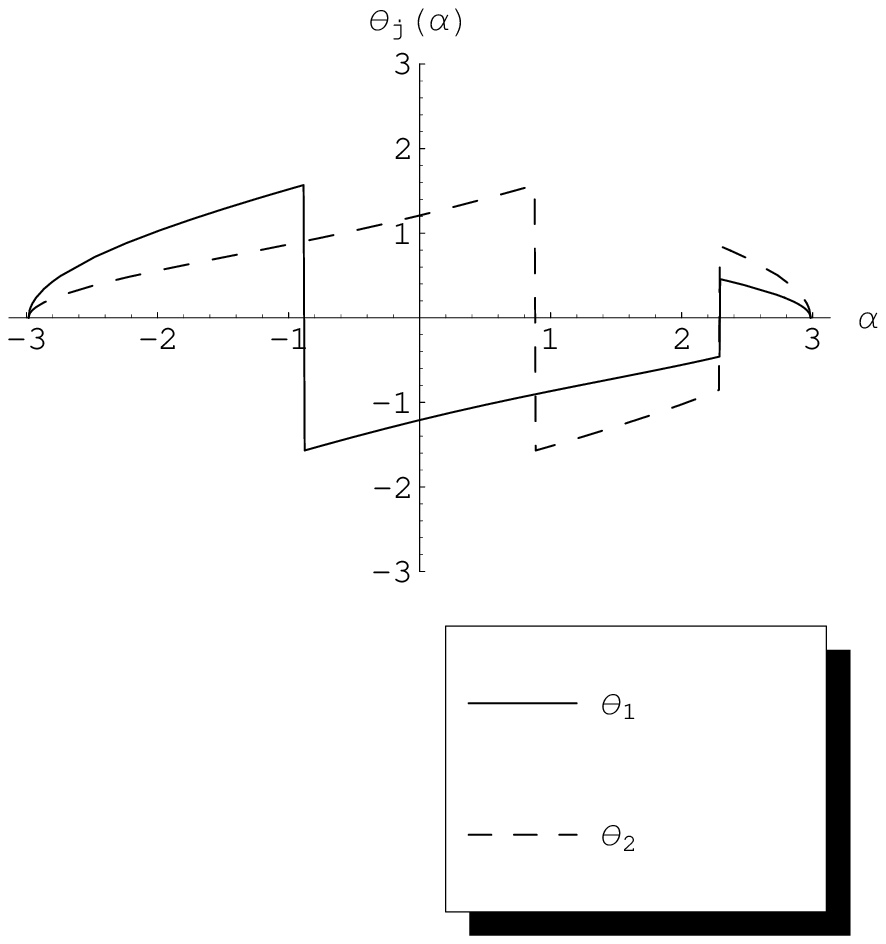}}
\hspace{0.2cm}
\fbox{\includegraphics[totalheight=6cm]{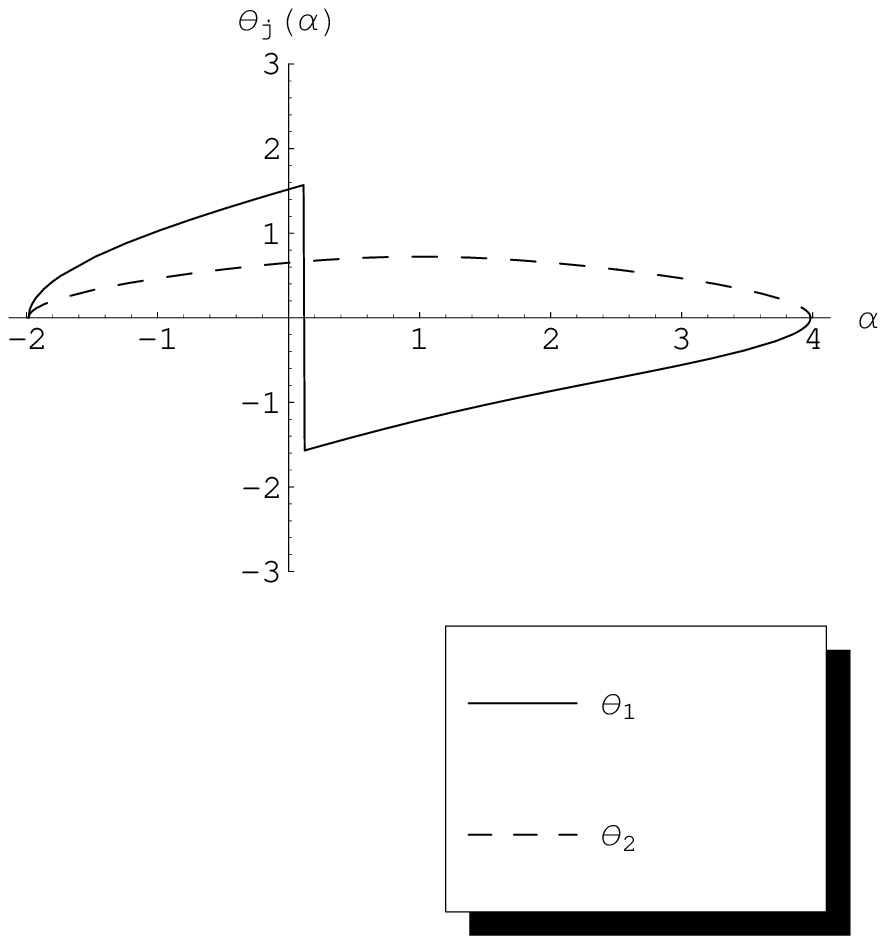}}
\caption{$k_1=-k_2=1$, $A_1=A_2=2$. On the left: $s_1=s_2=1$. On the right: $s_1=-s_2=1$. 
Here $c_1=1.5$, $c_2=1.6$}
\end{figure} 

Similarly, if there exist two gaps $\mathcal{I}^{(1)}_g$ and $\mathcal{I}^{(2)}_g$ we observe that the behaviour of the phases $\theta_1$ and $\theta_2$ as functions of $\alpha$ is different in the two gaps. An example of such dependence is displayed in Figure 5.

\begin{figure}[h!]\label{fig4}
\centering
\fbox{\includegraphics[totalheight=6cm]{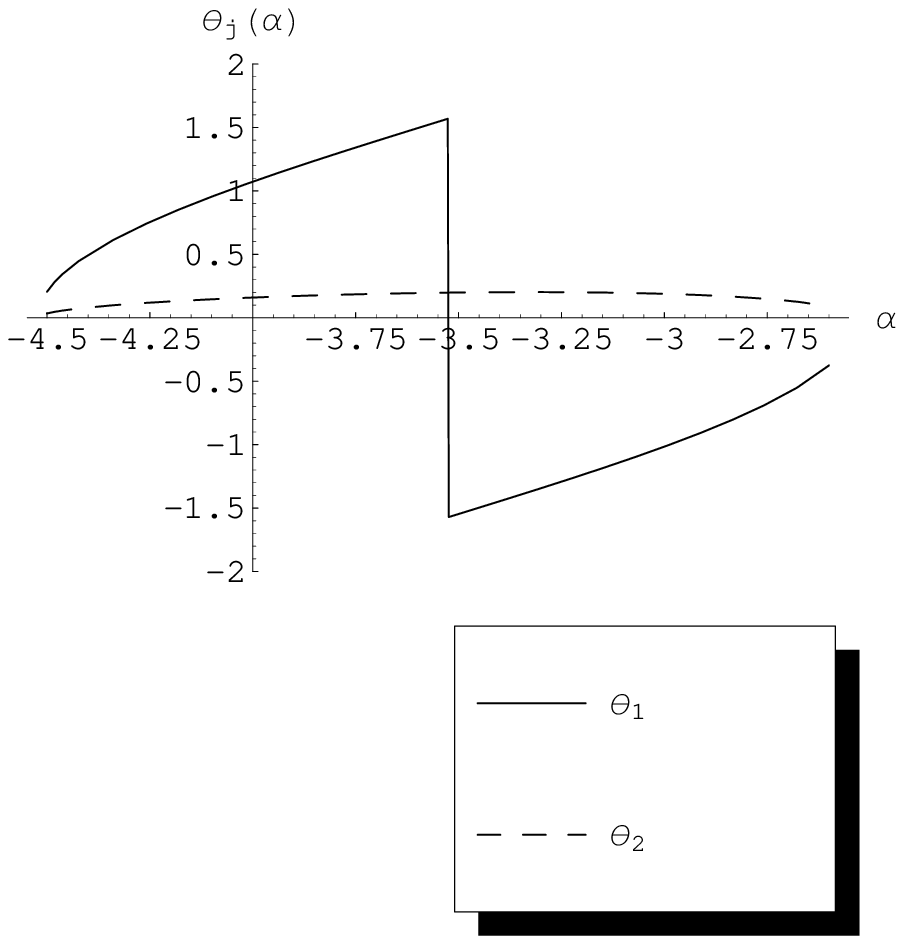}}
\hspace{0.2cm}
\fbox{\includegraphics[totalheight=6cm]{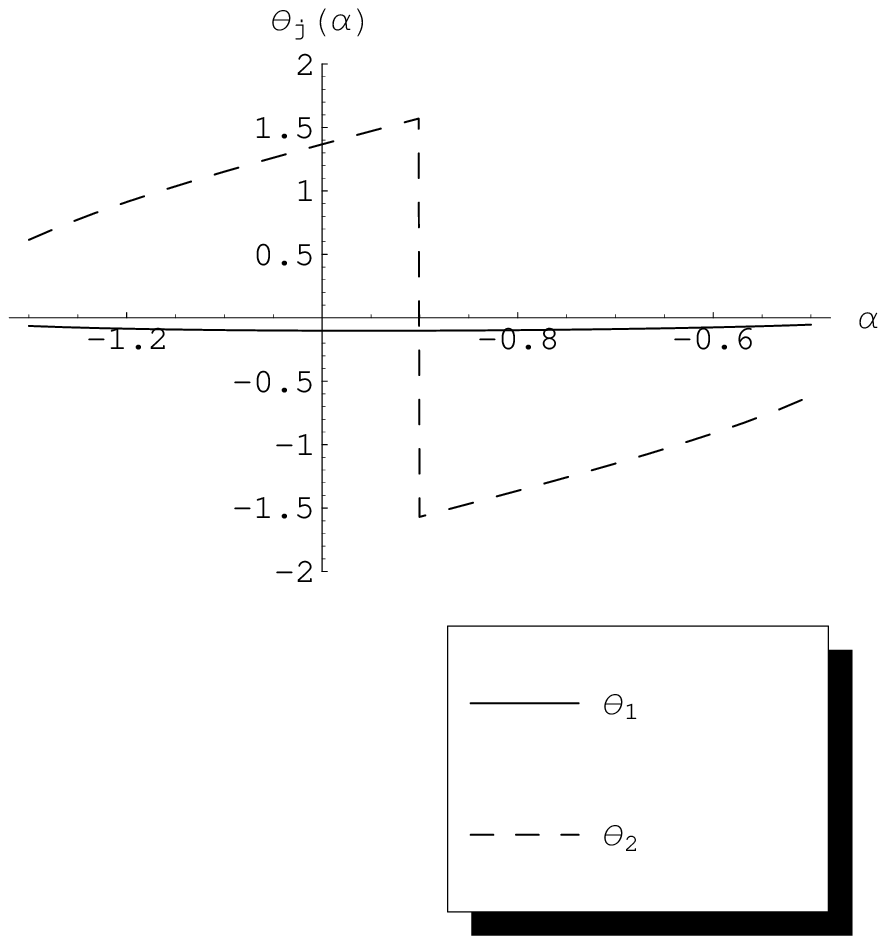}}
\caption{$s_1=s_2=1$, $k_1=7$, $k_2=2$, $A_1=1$, $A_2=1/2$;  $k^{(1)}_{-}=-4.5$, $k^{(1)}_{+}=-2.6$; $k^{(2)}_{-}=-1.5$, $k^{(2)}_{+}=-0.5$. On the left: $k^{(1)}_-\,<\,\alpha\,<\,k^{(1)}_+$. On the right: $k^{(2)}_-\,<\,\alpha\,<\,k^{(2)}_+$.}
\end{figure} 

%%%%%%%%%%%%%SEZIONE 5%%%%%%%%%%%%%%%%%%%%%%%%

\section{Conclusions and remarks}\label{conc}
The Zakharov--Shabat spectral problem, which leads to an important class of integrable partial differential equations, f.i. the Nonlinear Schr\"odinger, Sine--Gordon and modified Korteweg--de Vries equations, has been generalized to deal with matrix and vector dependent variables long time ago. This matrix generalization has been recently revisited in \cite{CD2004} and \cite{CD2006} and further generalized in order to derive integrable matrix partial differential equations with matrix, rather than scalar, coefficients. This generalization generates new integrable systems whose soliton dynamics may have peculiar properties which are known as boomeronic and are different from those of standard solitons.
In a previous paper \cite{ddt} we have adapted the standard Darboux dressing method to construct soliton solutions for a class of boomeronic--type equations. In this paper we have carried out this construction and obtained explicit expressions by dressing both the vacuum (i.e. vanishing) solution and the generic plane wave solution. We have specialized our formulae to the particularly interesting case of the resonant interaction of three waves, a well-known model which is of boomeronic--type. For this equation we have constructed a novel solution which describes three locked dark pulses. The stability of this solution, with respect to the sign of the coupling constants, is not investigated here and remains to be verified. Other potentially applicable integrable models are included in the present scheme. One of them, the Double three Wave Resonant Interaction (D3WRI) equation, seems particularly interesting in both fluid dynamics and nonlinear optics. In our present scheme, this  system of five partial differential equations takes the form
$$u_{1t}+c_1u_{1x}=-s_2w^*u_2\;\;,\;\;u_{2t}+c_2u_{2x}=s_1wu_1$$
 $$v_{1t}+c_1v_{1x}=-s_2w^*v_2\;\;,\;\;v_{2t}+c_2v_{2x}=s_1wv_1$$
 $$w_x=(c_2-c_1)(u_1^*u_2\,+\,sv_1^*v_2)$$
where $s_1=\pm 1\;,\;s_2=\pm1\;,\;s=\pm 1$ are just signs.
While these equations are obviously integrable, they may also follow, via the multiscale (alias slowly varying amplitude) approximation, from the resonant interaction of two triads of quasi--plane waves (with a commune mode), whose wave numbers $k_1\,,\,k_2\,,\,q_1\,,\,q_2\,,\,k_3$ and frequencies 
$\omega_1\,,\,\omega_2\,,\,\nu_1\,,\,\nu_2\,,\,\omega_3$ satisfy the resonance condition
$$k_2\,-\,k_1\,=\,q_2\,-\,q_1\,=\,k_3\;\;,\;\;\omega_2\,-\,\omega_1\,=\,\nu_2\,-\,\nu_1\,=\,\omega_3\;.$$

Regarding the 3WRI and the D3WRI equations, the following remark is appropriate. Since they are first order differential equations any nonsingular linear coordinate transformation of the $(x\,,\,t)$ plane in itself leaves these equations formally covariant. As a consequence, in different applications the coordinates $ x$ and $t$ may be given different physical meaning. Since we have constructed special solutions, this meaning is irrelevant in our present context. On the contrary, the meaning of these coordinates $ x$ and $t$ becomes relevant if one addresses the initial value problem, since one has to single out the \emph{evolution} variable. Indeed in our present dressing construction, the two linear equations of the Lax pair (\ref{lax}) have been treated on the same foot. Instead, the solution of the initial value problem requires that one of the two Lax equations plays the role of the spectral problem which provides the spectral Fourier--type data, while the other linear equation entails the evolution of such data. In this respect we point out that the initial value problem associated to the variable $t$ as  evolution variable (\emph{time} in the usual physical meaning) is still unsolved. This is due to the fact that the matrix $T(x,t,k)$, see (\ref{tLaxoperator}), depends asymptotically, as $x\rightarrow \pm \infty$,
on the unknown matrix $U$ through the asymptotic values of the matrices $W^{(\pm)}$, see the system (\ref{redboomeron}). Finding the solution of this open problem is left to future investigation.

%%%%%%%%%%%%%%%%%%%RINGRAZIAMENTI%%%%%%%%%%%%%%%
\textbf{Acknowledgements}
Of the two authors, one, S L, acknowledges financial support in first instance from EPSRC (EP/E044646/1) and then from NWO though the scheme VENI, and the other, A D, acknowledges financial support from the University of Roma ``Sapienza''. Part of this work has been carried out while both authors were visiting the Centro Internacional de Ciencias in Cuernavaca, Mexico, whose hospitality is gratefully acknowledged. 
%%%%%%%%%%%%%%%%%%%BIBLIOGRAFIA%%%%%%%%%%%%%%%%%%

 \end{document}